\documentclass[aps,prd,amsmath,floats,floatfix,twocolumn,nofootinbib,superscriptaddress,showpacs]{revtex4-1}
\usepackage{graphicx}
\usepackage{multirow}
\usepackage{color}
\usepackage{latexsym}
\usepackage{amsfonts}
\usepackage{amsmath}
\usepackage[colorlinks]{hyperref
}
\hypersetup{
  colorlinks=true,        % false: boxed links; true: colored links
  linkcolor=blue,         % color of internal links
  citecolor=cyan,         % color of links to bibliography
}
\graphicspath{{figures/}} % Location of the graphics files

\newcommand{\be}{\begin{eqnarray}}
\newcommand{\ee}{\end{eqnarray}}

\begin{document}
	
\title{Spinning test particle motion around a rotating wormhole}
	
    \author{Farrux Abdulxamidov}
	\email{farrukhabd63@gmail.com}%XXX@astrin.uz
    \affiliation{National University of Uzbekistan, Tashkent 100174, Uzbekistan}
    \affiliation{Ulugh Beg Astronomical Institute, Astronomy St. 33, Tashkent 100052, Uzbekistan}
	%%%%%%%%%%%%%%%%%%%%%%%%%%%%%%%%%%%%%%%%%%%%%%%%%%%%%%%%%
	\author{Carlos~A.~Benavides-Gallego}
	\email[Corresponding author: ]{cabenavidesg20@shao.ac.cn}
	\affiliation{Shanghai Astronomical Observatory, 80 Nandan Road, Shanghai 200030, P. R. China}	
	%%%%%%%%%%%%%%%%%%%%%%%%%%%%%%%%%%%%%%%%%%%%%%%%%%%%%%%%%
	\author{Wen-Biao Han}
	\email[]{wbhan@shao.ac.cn}
	\affiliation{Shanghai Astronomical Observatory, 80 Nandan Road, Shanghai 200030, P. R. China}
	\affiliation{Hangzhou Institute for Advanced Study, University of Chinese Academy of Sciences, Hangzhou 310124, China}
    \affiliation{School of Astronomy and Space Science, University of Chinese Academy of Sciences, Beijing 100049, China}
    \affiliation{Shanghai Frontiers Science Center for  Gravitational Wave Detection, 800 Dongchuan Road, Shanghai 200240, China}
	%%%%%%%%%%%%%%%%%%%%%%%%%%%%%%%%%%%%%%%%%%%%%%%%%%%%%%%%%%
	\author{Javlon~Rayimbaev}
	\email{javlon@astrin.uz}
	\affiliation{Ulugh Beg Astronomical Institute, Astronomy St. 33, Tashkent 100052, Uzbekistan}
	\affiliation{Akfa University, Kichik Halqa Yuli Street 17,  Tashkent 100095, Uzbekistan}
	\affiliation{National University of Uzbekistan, Tashkent 100174, Uzbekistan}
	\affiliation{Tashkent State Technical University, Tashkent 100095, Uzbekistan}
	%%%%%%%%%%%%%%%%%%%%%%%%%%%%%%%%%%%%%%%%%%%%%%%%%%%%%%%%%%
		\author{Ahmadjon Abdujabbarov}
	\email{ahmadjon@astrin.uz}
	\affiliation{Shanghai Astronomical Observatory, 80 Nandan Road, Shanghai 200030, P. R. China}
	%%%%	
	\affiliation{National University of Uzbekistan, Tashkent 100174, Uzbekistan} 
	%%%%
	
	\affiliation{Ulugh Beg Astronomical Institute, Astronomy St. 33, Tashkent 100052, Uzbekistan}
	%%%%
	\affiliation{Tashkent Institute of Irrigation and Agricultural Mechanization Engineers, Kori Niyoziy, 39, Tashkent 100000, Uzbekistan}
	\affiliation{Institute of Nuclear Physics, Tashkent 100214, Uzbekistan}

	\date{\today}
	
	\begin{abstract}
    In this work, we investigated the motion of spinning test particles around a rotating wormhole, extending, in this way, the previous work of Benavides-Gallego et al. in [Phys. Rev. D \textbf{101}, no.12, 124024] to the general case. Using the Mathisson-Papapetrous-Dixon equations, we study the effective potential, circular orbits, and the innermost stable circular orbit (ISCO) of spinning test particles. We found that both the particle and wormhole spins affect the location of the ISCO significantly. On the other hand, Similar to the non-rotating case, we also found two possible configurations in the effective potential: \textit{plus} and \textit{minus}. Furthermore, the minimum value of the effective potential is not at the throat due to its spin $a$, in contrast to the motion of the non-spinning test particles in a non-rotating wormhole, where the effective potential is symmetric, and its minimum value is at the throat of the wormhole. In the case of the ISCO, we found that it increases as the spin of the wormhole $a$ increases, in contrast to black holes where the presence of spin decreases the value of the ISCO. Finally, since the dynamical four-momentum and kinematical four-velocity of the spinning particle are not always parallel, we consider the superluminal bound, finding that the allowed values of $s$ change as the wormhole's spin $a$ increases.
	\end{abstract}
	
	\maketitle
	
%%%%%%%%%%%%%%%%%%%%%%%%%%%%%%%%%%%%%%%%%%%%%%%%%%%%%%%%%%%%%%%%%%%%%%%%%%%%%%%%%% Introduction %%%%%%%%%%%%%%%%%%%%%%%%%%%%
%%%%%%%%%%%%%%%%%%%%%%%%%%%%%%%%%%%%%%%%%%%%%%%%%%%%%%%%%%%%%%

    \section{Introduction\label{SecI}}
	 
    In 1935, Einstein and Rosen explored the possibility of an atomistic theory of matter and electricity without singularities~\cite{Einstein:1935tc}. The main idea behind their work was to consider the physical space as two identical sheets connected by a ``\textit{bridge}.'' In this geometrical representation, particles were the very bridges connecting these sheets. Hence, using the metric tensor $g_{\mu\nu}$ of general relativity and the fields $\varphi_\mu$ of electromagnetism, they modified the gravitational equations, demonstrating that it is possible to obtain regular spherically symmetric solutions. These solutions are currently known as Einstein-Rosen bridges or ``\textit{wormholes}''. 

    The representation of particles as ``\textit{bridges}'' could have been discovered back in 1916. According to G.~W.~Gibbons~\cite{G.W.Gibbons2015}, previous to Einstein-Rosen's paper, and a few months after the Schwarzschild solutions~\cite{Schwarzschild:1916uq, Schwarzschild:1916ae}, Ludwig Flamm submitted a manuscript in which he explores some geometrical aspects of both exterior and interior solutions of Schwarzschild space-time~\cite{L.Flamm1916}.  In the former case  (the exterior solution),  Flamm was able to show that ``\textit{the planar section is isometric to a surface of revolution, where the meridional curve is a parabola.}'' However, Flamm never contemplates the possibility of interpreting this result as a ``\textit{bridge}'' connecting to regions of space-time. For this reason, we assume that the work of Einstein and Rosen gave birth to the modern study of ``\textit{wormholes}''.

    Initially, the idea of ``\textit{bridges}'' was considered more attractive than that of black holes until Wheeler and Fuller showed that the Einstein-Rosen bridge\footnote{Also known as the Schwarzschild wormhole.} is unstable. Using a proper analysis, they discovered that the bridge would pinch-off in a finite time~\cite{Fuller:1962zza}. Therefore, Schwarzschild wormholes are not traversable. Nevertheless, the possibility of traversable wormholes was considered years latter in several works~\cite{Ellis:1973yv,Bronnikov:1973fh,Morris:1988tu, Morris:1988cz, Visser:1989kh, Visser:1989kg, Visser:1989vq, AzregAinou:1989wr, Visser:1989am, Poisson:1989zz, Visser:1990wj, Frolov:1990si, Visser:1990wi}. In particular, in Ref~\cite{Morris:1988cz}, Morris and Thorne used a different approach which allowed them to propose a set of ``\textit{basic wormhole criteria}.'' Thus, by assuming the wormhole geometry, they used the Einstein field equations to compute physical quantities, such as the total energy density, the tension per unit area, and the pressure, in terms of what they call the ``\textit{redshift}'' and ``\textit{shape}'' functions. Their analysis establishes three properties to describe a wormhole. Firstly, the space-time is assumed to be static and spherically symmetric. Secondly, any wormhole solution should contain a throat that connects two asymptotically flat regions of space-time. This property is deeply related to the ``\textit{shape function}''. Finally, the solution must be horizonless. Since wormholes causally connect two different portions of the space-time by the throat, the presence of a horizon would prevent the two universes from being connected causally. This condition is satisfied by demanding the ``\textit{redshift function}'' to be finite everywhere. 
    
    On the other hand, the most relevant conclusion in Morris and Thorne's analysis is the necessity of ``\textit{exotic matter}'' to generate a traversable wormhole. According to the authors, an observer passing through the throat with a radial velocity close to the speed of light would perceive negative energy. Therefore, from the classical point of view, traversable wormhole solutions violate the well-known \textit{energy conditions}, established precisely to avoid negative energy densities. However, from the quantum point of view, there are some situations in which such violations may be physically valid. The quantum mechanical creation of particles, for example~\cite{Martin-Moruno:2013sfa}. In this sense, one can not entirely rule out the possibility of the existence of the exotic material required for the throat of a traversable wormhole to hold.
    
    The idea of wormholes has been considered in different scenarios, such as Einstein’s gravity~\cite{Echeverria:1991nk,Deser:1992ts,Deser:1993jr,Hochberg:1997wp,Abdujabbarov:2016efm,Abdujabbarov:2009ad,Pugliese:2011xn,Willenborg:2018zsv,Arganaraz:2019fup,Stuchlik:2021tcn,Benavides-Gallego:2021lqn} and alternatives theories of gravity~\cite{Moffat:1991xp,Carlini:1992jda,Bhawal:1992sz,Letelier:1993cj,Vollick:1998qf}. In Ref.~\cite{Benavides-Gallego:2021lqn},  Benavides-Gallego et al. investigated the motion of spinning test particles around traversable wormholes. Using the Mathisson-Papapetrous-Dixon (MPD) equations, the authors computed the effective potential and showed that it is affected by the adimensional spin $s$ of the test particle. When $s=0$, the particle follows the geodesic equation, and its effective potential is symmetric: its behavior is the same in both the lower and upper universes. Therefore, the innermost stable circular orbit (ISCO) is at the same distance from the wormhole's throat. 
    
    On the other hand, when the spin $s\neq0$, there is a different kind of symmetry which depends on the sing of the particle's spin $s$ and $\mathcal{L}$, the particle's dimensionless angular momentum. In that case, there are two possible configurations which define two different symmetries: ``\textit{plus}'' and ``\textit{minus}''. In this sense, the effective potential has the same behavior if it has the same configuration. For example, a particle with negative $s$ and negative $\mathcal{L}$ will have the same behavior as a particle with both $s$ and $\mathcal{L}$ positive. In other words, the effective potential of a particle with $s$ parallel to the symmetry axis ($s>0$) and moving counterclockwise ($\mathcal{L}>0$) is the same as that of a particle with $s$ antiparallel to the symmetry axis ($s<0$) and moving clockwise around the wormhole ($\mathcal{L}<0$), see Fig.~3 of Ref.~\cite{Benavides-Gallego:2021lqn}. The authors also found a ``\textit{mirror behavior}'' when the ``\textit{plus}'' and ``\textit{minus}'' configurations are considered at the same time. To explain this behavior, the authors consider two spinning particles moving counterclockwise with the same angular momentum $\mathcal{L}$, one in the lower universe (with $s>0$) and the other in the upper universe (with $s<0$). In this case, the effective potential has the same behavior for each particle, resembling the case of a non-spinning test particle, see the first panel in figure 3 of Ref.~\cite{Benavides-Gallego:2021lqn}.  
    
    The existence of two configurations affects the location of the ISCO. As mentioned before, in the case of non-spinning test particles, one finds a single value for $l_{ISCO}$ located at the same distance from the wormhole's throat in both universes\footnote{$l$ is the radial coordinate in Ref.~\cite{Benavides-Gallego:2021lqn}. The value $l=0$ represents the wormhole's throat.}. However, the behavior is different when we consider spinning test particles. If $|s|\geq 1$, one finds only one possible value for the ISCO. On the other hand, if $s$ belongs to the interval $-1<s<1$, one encounters two possible values for the ISCO. One of these values is always closer to the wormhole's throat.   
    
    One significant conclusion from Ref.~\cite{Benavides-Gallego:2021lqn} is the constrain obtained for the particle's spin $s$. It is well-known that the dynamical four-momentum $p^\alpha$ and the kinematical four-velocity $u^\alpha$ of a spinning test particle are not always parallel. In this sense, although  $p_\alpha p^\alpha = -m^2$ holds, the normalization $u_\alpha u^\alpha = -1$ does not. Therefore, while the spinning particle moves closer to the center of symmetry, $u^\alpha$ increases, and eventually, for certain values of the spin $s$ and radius $l$, some components of the four-velocity may diverge. This means that for a certain value of $s$, the particle's trajectory changes from time-like to space-like, becoming, in this way, superluminal. From the physical point of view, the space-like motion does not have any meaning because the transition to $u_\alpha u^\alpha > 0$ is not allowed for real particles. Therefore, one must impose an additional constrain defined by the relation $u_\alpha u^\alpha = 0$, the superluminal bound. In the case of a non-rotating wormhole, spinning test particles will always move in the time-like region as long as $|s|<1.5$~\cite{Benavides-Gallego:2021lqn}.
    
    In this paper, we consider the motion of spinning test particles around a rotating wormhole, generalizing in this way the results obtained by Benavides-Gallego et al. in Ref.~\cite{Benavides-Gallego:2021lqn}. We organize our work as follows. In Sec.~\ref{SecII}, we review the properties of rotating wormholes discussed in Ref.~\cite{Teo:1998dp} by E.~Teo and its connection with the ideas of traversable wormholes described in Ref.~\cite{Morris:1988cz}. Next, in Sec.~\ref{SecIII}, we discuss the motion of spinning particles in a static and axially symmetric space-time, where we use the MPD equations to obtain the effective potential and the superluminal condition. Then, in Sec.~\ref{SecIV}, we apply the results of Sec.~\ref{SecIII} to Teo's wormhole solution. We compute the effective potential, circular orbits, and the innermost stable circular orbit (ISCO). We also use the superluminal bound to find a constraint for the allowed spin of the particle. Finally, in Sec.~\ref{SecV}, we summarize our work and discuss the implications of the results. Throughout the manuscript, we use geometrized units setting $G=c=1$ and $b_0=M=1$.

%%%%%%%%%%%%%%%%%%%%%%%%%%%%%%%%%%%%%%%%%%%%%%%%%%%%%%%%%%%%%%%%%%%%%%%%%%%%%%%%%%%%% Section II %%%%%%%%%%%%%%%%%%%%%%%%%%%%
%%%%%%%%%%%%%%%%%%%%%%%%%%%%%%%%%%%%%%%%%%%%%%%%%%%%%%%%%%%%%%
	
	\section{Stationary, axisymmetric space-times\label{SecII}}
	
	In Ref.~\cite{Teo:1998dp}, Teo follows the same paradigm as Morris-Thorne in Ref.~\cite{Morris:1988cz}, i.e. he first assumes the space-time geometry and then uses the Einstein field equations to deduce the form of the matter required to maintain the wormhole. Hence, the author begins by considering a stationary and axially symmetric space-time.

    It is well-known that space-times are said to be stationary if it possesses a time-like Killing vector field $\xi^\alpha\equiv(\partial/\partial t)^\alpha$, which generates invariant time translations. On the other hand, a space-time is axisymmetric if it possesses a space-like killing vector field $\psi^\alpha\equiv (\partial/\partial \varphi)^\alpha$, related to invariant rotations with respect to $\varphi$. Therefore, a space-time is stationary and axisymmetric if it possesses both $\xi^\alpha$ and $\psi^\alpha$ killing vector fields satisfying the following commutation relation~\cite{Wald:1984rg},
	\begin{equation}
	    \label{s2ae1}
	    [\xi,\psi]=0.
	\end{equation}
	The commutativity of $\xi^\alpha$ and $\psi^\alpha$ in Eq.~(\ref{s2ae1}) allow us to choose a coordinate system in such a way that the Killing vectors represent the directions in which the space-time has symmetries~\cite{Wald:1984rg}. Hence, one can set $x^0=t$, $x^1=\varphi$, $x^2$, and $x^3$ as a coordinate system. The stationary and axisymmetric character of the space-time requires the metric components to be independent of $t$ and $\varphi$. Therefore, the metric takes the form,
	\begin{equation}
	    \label{s2ae2}
	    ds^2=g_{\mu\nu}(x^2,x^3)dx^\mu dx^\nu.
	\end{equation}
    From the physical point of view, stationary and axisymmetric space-times have been of considerable interest in the study of black holes and stars since this kind of geometry describes the exterior gravitational field of rotating bodies, see Refs~\cite{Hartle:1967he, Hartle:1968si, Thorne:1971R} and references therein. 
	
	Thorne discusses the properties of static and axisymmetric space-times in Ref.~\cite{Thorne:1971R}. There, he stars by pointing out that $(t,\varphi,x^2,x^3)$ and $(t,\varphi +2\pi,x^2,x^3)$ represent the same point. This behavior is due to the fact that $\varphi$ is an angular coordinate about the rotation axis. As a consequence, $\varphi$ belongs to the interval $[0,2\pi)$. 
	
	On the other hand, because the space-time is axially symmetric, it must be invariant under a simultaneous inversion of $t$ and $\varphi$, i.e. the space-time does not change if $t\rightarrow -t$ and $\varphi\rightarrow -\varphi$. As a consequence, the metric coefficients $g_{t2}$, $g_{t3}$, $g_{\varphi 2}$ and $g_{\varphi 3}$ must vanish because they change the sign under simultaneous inversion of $\varphi$ and $t$. Therefore, the line element of Eq.~(\ref{s2ae2}) simplifies even more and reduces to~\cite{Papapetrou:1966zz,Carter:1969zz}
	\begin{equation}
	    \label{s2ea3}	    ds^2=g_{00}dt^2+2g_{01}dtd\varphi+g_{11}d\varphi^2+g_{ij}dx^idx^j,
	\end{equation} 
    with $i,j=2,3$. Here, the presence of the term $g_{01}$ is related to the well-know \textit{dragging effect}, see Appendix~\ref{appenA}. 
    
    The coordinates in Eq.~(\ref{s2ea3}) are uniquely determine up to a coordinate transformation of the form~\cite{Thorne:1971R}
    \begin{equation}
    \label{s2ea4}
	    \begin{array}{ccc}
	    \overline{x}^2=\overline{x}^2(x^2,x^3)&\text{and}&            \overline{x}^3=\overline{x}^3(x^2,x^3).	    
    \end{array}
    \end{equation}
    The freedom in such transformations can be used to simply the mathematics in the Einstein's field equations or adapt the geometry to specific problems. Under transformation of the form given in Eq.~(\ref{s2ea4}), the components $g_{00}$, $g_{01}$, and $g_{11}$ are invariant. Hence, $g_{00}=\xi_\alpha\xi^\alpha$, $g_{01}=\psi_\alpha\xi^\alpha$, and $g_{11}=\psi_\alpha\psi^\alpha$~\cite{Wald:1984rg,Thorne:1971R}.
	
	Finally, the space-time described in Eq.~(\ref{s2ea3}) must be asymptotically flat. This means that $g_{00}\rightarrow 1$, $g_{01}\rightarrow 1/r$ and $g_{11}\rightarrow r^2\sin^2\theta$ as $r\rightarrow \infty$. The asymptotically-flatness of the line element~(\ref{s2ea3}) is important to define the mass, and the angular momentum\footnote{Here $\theta$ and $r$ are the usual spherical coordinates, but not necessarily the same as $x^2$ and $x^3$\cite{Thorne:1971R}}.

    \subsection{Canonical form of the rotating wormhole space-time}

	In the particular case of the rotating wormhole solution, by considering $g_{22}=g_{33}=g_{11}/\sin^2 x^2$ and $g_{23}=0$, the line element of Eq.~(\ref{s2ea3}) can be expressed as~\cite{Teo:1998dp}
    \begin{equation}
    \label{s2be1}
    ds^2=-N^2dt^2+e^\mu dr^2+r^2K^2\left[d\theta^2+\sin^2\theta(d\varphi-\omega dt)^2\right],
    \end{equation}
    where the functions\footnote{Also known as the four \textit{gravitational potentials}.} $N$, $\mu$, $K$ and $\omega$ only depend on the $(x^2,x^3)\equiv (\theta,r)$. Following Ref.~\cite{Chandrasekhar:1985kt}, it is possible to show the dragging effect in the space-time described by the line element in Eq.~(\ref{s2be1}), see Appendix~\ref{appenA}.
	
	Once the line element of a stationary and axisymmetric space-time is defined, Teo discusses some important features related to the metric. For example, he points out that the function $K(r,\theta)$ is a positive, nondecreasing function of $r$, which he uses to define the ``\textit{proper distance}'' $R\equiv r K(r,\theta)$ (with $\partial R/\partial r >0$) measure at $(r,\theta)$ from the origin. In this sense, one can interpret the value $2\pi R\sin\theta$ as the proper circumference of the circle located at the point with coordinates $(r,\theta)$.
	
	On the other hand, the metric defined in Eq.~(\ref{s2be1}) has the discriminant\footnote{From now on, we associate the coordinates $t$, $\phi$, $r$, $\theta$ with the numbers, 0, 1, 2 and 3, respectively.}~\cite{Teo:1998dp},
    \begin{equation}
    \label{s2be2}
    D^2=g^2_{t\varphi}-g_{tt}g_{\varphi\varphi}=(N(r,\theta)K(r,\theta)\sin\theta)^2.
    \end{equation}
    According to Teo, the existence of horizons is determined by the function $N(r,\theta)$, which plays the role of the ``\textit{redshift function}''. Whenever, $N=0$, $D^2=0$, implying the presence of an event horizon. Therefore, to avoid a singular behavior of the metric ($D^2\neq0$) on the rotation axis $\theta=0$ and $\theta=\pi$, Teo imposes the regularity conditions on \textit{the gravitational potential}. These conditions state that the partial derivatives with respect to $\theta$ of $N(r,\theta)$, $\mu(r,\theta)$, and $K(r,\theta)$ must vanish on the rotation axis. The regularity conditions are, in this sense, crucial to establish the wormhole geometry in the line element of Eq.~(\ref{s2be1}) since wormholes, according to Ref.~\cite{Morris:1988cz}, must be horizonless.
	
	In analogy to Morris-Thorne function $\mu$, Teo defines a similar function by including the dependence on $\theta$. Therefore, $\mu(r,\theta)$ is given by~\cite{Teo:1998dp}
    \begin{equation}
    \label{s2be3}
    \mu(r,\theta)=-\ln\left(1-\frac{b(r,\theta)}{r}\right);
    \end{equation}
    with $b(r,\theta)$ playing the role of the ``\textit{shape function}''. In this sense, the radial coordinate must be constrained to $r\geq b$, where the throat is located at $r=b$. In this way, Eq.~(\ref{s2be1}) reduces to the Morris-Thorne case when there is no rotation, e. i. $N(r,\theta)\rightarrow e^\Phi(r)$, $b(r,\theta)\rightarrow b(r)$, $K(r,\theta)\rightarrow1$ and $\omega(r,\theta)\rightarrow0$. Moreover, Teo assumes the \textit{gravitational potentials} to be well-behaved at the wormhole's throat. The reason for such an assumption has to do with the singularity-free behavior at the throat. If one computes the curvature scalar of the metric in Eq.~(\ref{s2be1}) (evaluated at the throat), it is possible to see that it has the form~\cite{Harko:2009xf}
    \begin{equation}
    \label{s2be4}
    \begin{aligned}
    \mathcal{R}&=-\frac{1}{r^2 K^2}\left(\mu_{\theta\theta}+\frac{1}{2}\mu^2_\theta\right)-\frac{\mu_\theta}{Nr^2K^2}\frac{(N\sin\theta)_\theta}{\sin\theta}\\
    &-\frac{2}{Nr^2K^2}\frac{(N_\theta \sin\theta)_\theta}{\sin\theta}-\frac{2}{r^2K^3}\frac{(K_\theta\sin\theta)_\theta}{\sin\theta}\\
    &e^{-\mu}\mu_r[\ln(N r^2K^2)]_r+\frac{\sin^2\theta\omega^2_\theta}{2N^2}+\frac{2}{r^2K^4}(K^2+K^2_\theta),
    \end{aligned}
    \end{equation}
    where the subscripts denote partial derivatives with respect to $\theta$ and $r$. From the last equation, one can see that $\mathcal{R}$ could have a singular behavior due to terms~\cite{Teo:1998dp,Harko:2009xf} 
    \begin{equation}
    \label{s2be5}
    \begin{aligned}
    \mu_{\theta\theta}+\frac{1}{2}\mu^2_\theta&=\frac{b_{\theta\theta}}{r-b}+\frac{3}{2}\frac{b^2_\theta}{(r-b)^2},\\
    \mu_\theta&=\frac{b_\theta}{r-b}.
    \end{aligned}
    \end{equation}
    Therefore, to avoid singularities in the curvature scalar at the throat, it is necessary that $b_\theta=b_{\theta\theta}=0$. Ergo, the throat is at a constant value of $r$. 
   
   As mentioned before, if one wants the line element in Eq.~(\ref{s2be1}) to describe a wormhole, it is crucial to satisfy the so-called ``\textit{flare-out}'' condition at the throat. Following the same process described in Ref.~\cite{Morris:1988cz}, Teo embeds the space-time in a higher-dimensional space by considering a constant value of $\theta$ in a slice of constant $t$, what Morris-Thorne think of as a picture of the whole space-time at a fixed moment $t$. 

    After embedding the metric, Teo found the following ``\textit{flare-out}'' condition at the throat~\cite{Teo:1998dp}
    \begin{equation}
    \label{s2be6}
    \frac{d^2r}{dz^2}=\frac{b-b_r r}{2b^2}>0,
    \end{equation}
    which is the same condition as in the Morris-Throne wormhole~\cite{Morris:1988cz}. Since $b_\theta=0$, it is possible to define a new radial coordinate $l^2=r^2+b^2$ in the vicinity of the throat, satisfying the relation
    \begin{equation}
    \label{s2be7}
    \frac{dl}{dr}\equiv\pm\left(1-\frac{b}{r}\right)^{-1/2}.
    \end{equation}
    Hence, in the immediate vicinity of the throat\footnote{To first order in $r-r_0$, with $r_0$ the location of the throat.} the line element in Eq.~(\ref{s2ae1}), reduces to~\cite{Teo:1998dp}
    \begin{equation}
    \label{s2be8}
    \begin{aligned}
    ds^2&=-N^2(l,\theta)dt^2+dl^2+r^2(l)K^2(l,\theta)\\
    &\times[d\theta^2+\sin^2\theta(d\varphi-\omega(l,\theta)dt)^2].
    \end{aligned}
    \end{equation}
    The metric expressed in this way smoothly connects two asymptotic regions of the space-time across the throat, in contrast to Eq.~(\ref{s2ae1}), where the radial coordinate $r$ is singular. If the \textit{shape function} does not depend on $\theta$, Eq.~(\ref{s2be7}) is valid everywhere and the coordinate $l$ takes the range $(-\infty,\infty)$. As a consequence, the metric in Eq.~(\ref{s2be7}) covers the whole space-time, and we can assume the wormhole throat is at $l=0$ and define the upper universe when $l>0$, and the lower universe when $l<0$.  
    
	According to Morris-Thorne, one can use the space-time in Eq.~(\ref{s2ae1}) to compute the non-vanishing components of the stress-energy tensor. To do so, one needs to consider a ``\textit{local Lorentz frame}'', where physical ``observations'' are performed by a local observer, who remains at rest with respect to the coordinate system $(t,\theta,r,\varphi)$. In this frame, the components are~\footnote{Here we use the same notation as in Ref.~\cite{Chandrasekhar:1985kt}.} $T_{(t)(t)}$, $T_{(t)(\varphi)}$, $T_{(\varphi)(\varphi)}$, and $T_{(i)(j)}$. These components have the usual physical meaning. For example, $T_{(t)(t)}$ is the mass-energy density, while $T_{(t)(\varphi)}$ represents the rotation of the matter distribution. In Ref.~\cite{Harko:2009xf}, Harko et al. obtained the expressions evaluated at the throat.

	Finally, using the null energy condition~\cite{Wald:1984rg}
    \begin{equation}
    \label{s2be9}
    R_{\alpha\beta}\kappa^\alpha\kappa^\beta\geq 0,
    \end{equation}
    where $R_{\alpha\beta}$ is the Ricci tensor and $\kappa^\alpha$ a null vector given by~\cite{Teo:1998dp} 
    \begin{equation}
    \label{s2be10}
    \kappa^\alpha=\left(\frac{1}{N},-e^{-\mu/2},0,\frac{\omega}{N}\right),
    \end{equation}
    Teo have found that~\cite{Teo:1998dp} 
    \begin{equation}
    \label{s2be11}
    \begin{aligned}
    R_{\alpha\beta}\kappa^\alpha\kappa^\beta&=e^{-\mu}\mu_r\frac{(rK)_r}{rK}-\frac{\omega^2_\theta\sin^2\theta}{2N^2}-\frac{1}{4}\frac{\mu^2_\theta}{(rK)^2}\\
    &-\frac{1}{2}\frac{(\mu_\theta\sin\theta)_\theta}{(rK)^2\sin\theta}+\frac{(N_\theta\sin\theta)_\theta}{(rK)^2N\sin\theta}<0.
    \end{aligned}
    \end{equation}
    Nevertheless, He remarks that by choosing $N$, and $\mu$ appropriately, $R_{\alpha\beta}$ could be positive at some point in the interval $(0,\pi)$. Consequently, it is possible to move the exotic matter that supports the wormhole around the throat so that an infalling observer would not encounter it. One example of such space-time is given by considering~\cite{Teo:1998dp} 
    \begin{equation}
    \label{s2be12}
    \begin{aligned}
    N&=K=1+\frac{(4J\cos\theta)^2}{r},\\
    b&=\frac{b^2_0}{r},\\
    \mu&=-\ln \left(1-\frac{b}{r}\right)\\
    \omega&=\frac{2J}{r^3}.
    \end{aligned}
    \end{equation}
   
%%%%%%%%%%%%%%%%%%%%%%%%%%%%%%%%%%%%%%%%%%%%%%%%%%%%%%%%%%%%%%%%%%%%%%%%%%%%%%%%%%% Section III %%%%%%%%%%%%%%%%%%%%%%%%%%%%
%%%%%%%%%%%%%%%%%%%%%%%%%%%%%%%%%%%%%%%%%%%%%%%%%%%%%%%%%%%%%%   
   	\section{Equations of motion \label{SecIII}}

    In this section, we review the theoretical background necessary to investigate the motion of spinning test particles. It is well-known that this problem was considered for the first time by Mathisson in 1937 when he studied the problem of extended bodies in general relativity (GR). According to Mathisson, the motion of spinning test particles does not follow the usual geodesic equation of GR due to the coupling between the Riemann curvature tensor and the spin of the moving particle~\cite{Mathisson:1937zz}. Papapetrou considers the same problem in Refs.~\cite{Papapetrou:1951pa, Corinaldesi:1951pb}, where he established a similar approach. Later, Tulczyjew improved on the methods of Mathisson~\cite{tulczyjew1959motion, BWTulzcyjew1962} while Moller and others made improvements in the definition of center-of-mass~\cite{moller1949definition,beiglbock1967center,dixon1964covariant, Dixon:1970zza, Dixon:1970zz, ehlers1977dynamics}. Today, the equations that describe the motion of extended bodies with spin and mass are known as the Mathisson-Papapetrous-Dixon (MPD) equations. Recently, some authors modified the MPD equations, see Refs.~\cite{Deriglazov:2017jub, Deriglazov:2018vwa}.

	Mathematically, the MPD equations are given by
    \begin{equation}
    \label{s3e1}
    \begin{aligned}
    \frac{Dp^\alpha}{d\lambda}&=-\frac{1}{2}R^\alpha_{\;\;\beta\delta\sigma}u^\beta S^{\delta\sigma},\\
    \frac{D S^{\alpha\beta}}{d\lambda}&=p^\alpha u^\beta-p^\beta     u^\alpha,
    \end{aligned}
    \end{equation}
    where $D/d\lambda\equiv u^\alpha\nabla_\alpha$ is the projection of the covariant derivative along the particle's trajectory, $u^\mu=dx^\mu/d\lambda$ is the 4-velocity of the test particle, $p^\alpha$ is the canonical 4-momentum, $R^\alpha_{\;\;\beta\delta\sigma}$ is the Riemann curvature tensor, and $\lambda$ is an affine parameter. The second rank tensor $S^{\alpha\beta}$ is antisymmetric, $S^{\alpha\beta}=-S^{\beta\alpha}$. 
	
	Expressed in that form, Eq.~(\ref{s3e1}) shows us the coupling between the Riemann tensor $R^\alpha_{\;\;\beta\delta\sigma}$ and the spin of the moving particle. To see this coupling clearly, let's consider the well-known geodesic equation of GR
    \begin{equation}
    \label{s3e2}
    u^\beta\partial_\beta u^\alpha             +\Gamma^\alpha_{\;\;\;\sigma\beta}u^\sigma u^\beta=0.
    \end{equation}
    In terms of the particle 4-momentum and the projection of the covariant derivative along the particle's trajectory, Eq.~(\ref{s3e2}) reduces to 
    \begin{equation}
    \label{s3e3}
    \frac{Dp^\alpha}{d\lambda}=0.
    \end{equation} 
    Therefore, comparing Eqs.~(\ref{s3e1}) and (\ref{s3e3}), one can see how the interaction between the Riemann curvature tensor and the antisymmetric tensor $S^{\alpha\beta}$ does affect the motion of spinning test particles in curved space-times. 
	
	A crucial aspect in the MPD equations is related to the center of mass of the spinning test particle. In this sense, to solve the system in Eq.~(\ref{s3e1}), one needs to fix its center of mass. This is done by including the condition~\cite{tulczyjew1959motion,Saijo:1998mn}
    \begin{equation}
    \label{s3e6}
    S^{\alpha\beta}p_\alpha=0.
    \end{equation} 
    This condition is known as the Tulczyjew Spin Supplementary Condition (SSC)~\cite{Saijo:1998mn}. From Eq.~(\ref{s3e6}), the canonical momentum and the spin of the particle provide two independent conserved quantities given by the relations
    \begin{equation}
    \label{s3e7}
    \begin{aligned}
    p^\alpha p_\alpha&=-m^2,\\
    S^{\alpha\beta} S_{\alpha\beta}&=2S^2
    \end{aligned}
    \end{equation}
    Nevertheless, in contrast to the spinning test particle's canonical momentum conservation, it is important to point out that the squared velocity does not necessarily satisfy the condition
    \begin{equation}
    \label{s3e8}
    u_\alpha u^\alpha=-1,
    \end{equation}
    because the 4-vectors $p^\alpha$ and $u^\alpha$ are not always parallel. In this sense, to ensure that the particle's 4-velocity is always smaller than the speed of light, one needs to impose an additional condition: \textit{the superluminal bound}, see Sec.~\ref{SecIIIB}.

	Additionally, to the conserved quantities resulting from the Tulczyjew-SSC condition, one also has the conserved quantities associated to the space-time symmetries. As remarked in Sec.\ref{SecII}, the line element in Eq.~(\ref{s2be1}) has associated two Killing vector fields. One generating invariant time translations ($\xi^\alpha$) and the other related to invariant rotations with respect to $\varphi$ ($\psi^\alpha$). Hence, the conserved quantities associated to them can be obtained from the following relation, 
    \begin{equation}
    \label{s3e9}
    p^\alpha \kappa_\alpha-\frac{1}{2}S^{\alpha\beta}\nabla_\beta\kappa_\alpha=p^\alpha \kappa_\alpha-\frac{1}{2}S^{\alpha\beta}\partial_\beta\kappa_\alpha=\text{constant},
    \end{equation} 
    where $\kappa^\alpha$ is the Killing vector field.  In the last expression, we used the fact that the term $S^{\alpha\beta}\Gamma^\gamma_{\;\;\beta\alpha}$ in the covariant derivative $S^{\alpha\beta}\nabla_\beta\kappa_\alpha$ vanishes because $S^{\alpha\beta}$ is antisymmetric while  $\Gamma^\gamma_{\;\;\beta\alpha}$ is symmetric.
	
    \subsection{The effective potential\label{SecIIIA}}
   
    In the case of static axially symmetric space-times, Toshmatov and Malafarina obtained the most general expression for $V_{\text{eff}}$ in terms of the metric components $g_{\mu\nu}$ and its derivatives\footnote{Where ${}'$ denotes the derivative with respect to the radial coordinate} $g'_{\mu\nu}$ Ref.~\cite{Toshmatov:2019bda}. Here, we review and discuss the most important steps in the calculation.
   
    Let's start considering the line element of a stationary and axisymmetric space-time. As mentioned Sec.~\ref{SecII}, Eq.~(\ref{s2ea3}) give us its general form
    \begin{equation}
    \label{s3ae1}  ds^2=g_{tt}dt^2+g_{rr}dr^2+2g_{t\varphi}dtd\varphi+g_{\theta\theta}d\theta^2+g_{\varphi\varphi}d\varphi^2.
    \end{equation}
    As we pointed out above, due to the Killing vectors $\xi^\alpha=\delta^\alpha_t$ and $\psi^\alpha=\delta^\alpha_\varphi$, this space-time has associated two constants of motion: the energy $E$ and the angular momentum $L$. Hence, after using Eq.~(\ref{s3e9}), we obtain\footnote{There is a typo in the second relation of Eq.~(11) of Ref.~\cite{Toshmatov:2019bda}. It should be $S^{\alpha\beta}$.}
    \begin{eqnarray}
        \label{s3ae2}
    -E&=&p_t-\frac{1}{2}g_{t\alpha,\beta}S^{\alpha \beta}\\ \nonumber&=&p_t-\frac{1}{2}\left(g_{tt,r}S^{tr}+g_{\varphi t,r}S^{\varphi r}\right),\\
    L&=&p_{\varphi}-\frac{1}{2}g_{t\alpha,\beta}S^{\alpha\beta}\\ \nonumber&=&p_\varphi-\frac{1}{2}\left(g_{t\varphi,r}S^{tr}+g_{\varphi\varphi,r}S^{\phi r}\right).
        \end{eqnarray}
    In the last system of equations, we assume that the particle's motion is constrained to the equatorial plane ($\theta =\pi/2$). Due to this assumption, the metric functions depend only on the radial coordinate and $p_\theta= 0$. Furthermore, since $S^{\theta\alpha}=0$, the number of independent components of the spin tensor is reduced to three, i.e. $S^{tr}$, $S^{t\varphi}$, and $S^{r\varphi}$.  
   
    To solve the system in Eq.~(\ref{s3ae2}), it is necessary to express the components $S^{t\varphi}$ and  $S^{r\varphi}$ in terms of $S^{tr}$. To do so, one uses the Tulczyjew-SSC contidion (\ref{s3e6}), from which one obtains 
    \begin{equation}
    \label{s3ae3}
    \begin{aligned}
    S^{t\varphi}&=\frac{p_r}{p_{t}}S^{\varphi r}=-\frac{p_r}{p_{\varphi}}S^{tr}\\\\
    S^{r\varphi}&=-\frac{p_t}{p_{\varphi}}S^{rt}=\frac{p_t}{p_{\varphi}}S^{tr}.
    \end{aligned}
    \end{equation}
    Using these relations, Eq.~(\ref{s3ae2}) takes the form 
    \begin{equation}
    \label{s3ae4}
    \begin{aligned}
    -E&=p_t-\frac{1}{2}\left(g'_{tt}p_\varphi-g'_{\varphi t}p_t\right)\frac{S^{tr}}{p_\varphi}\\\\
    L&=p_\varphi-\frac{1}{2}\left(g'_{t\varphi}p_\varphi-g'_{\varphi\varphi}p_t\right)\frac{S^{tr}}{p_\varphi}.
    \end{aligned}
    \end{equation}
    Now, from the spin coservation and the normalization conditions in Eq.~(\ref{s3e7}), we have that 
    \begin{equation}
    \label{s3ae5}
    S^{tr}=\frac{p_\varphi s}{\sqrt{g_{rr}(g^2_{t\varphi}-g_{\varphi\varphi}g_{tt})}}.
    \end{equation}
    Here, $s=S/m$ represents the specific angular momentum of the particle, which can be positive or negative with respect to the direction of $p_\varphi$. After replacing Eq.~(\ref{s3ae5}) into Eq.~(\ref{s3ae4}) and solving the system for $p_t$ and $p_\varphi$, one obtains~\cite{Toshmatov:2019bda}
    \begin{equation}
    \label{s3ae6}
    \begin{aligned}
    p_t&=\frac{-E+s(AL+B E)}{1-Ds^2}\\\\
    p_\varphi&=\frac{L+s(B L +C E)}{1-Ds^2}.
    \end{aligned}
    \end{equation}
    with~\cite{Toshmatov:2019bda}
    \begin{equation}
    \label{s3ae7}
    \begin{aligned}
    A&=\frac{g'_{tt}}{2\sqrt{g_{rr}(g^2_{t\varphi}-g_{\varphi\varphi}g_{tt})}},\\\\
    B&=\frac{g'_{t\varphi}}{2\sqrt{g_{rr}(g^2_{t\varphi}-g_{\varphi\varphi}g_{tt})}},\\\\
    C&=\frac{g'_{\varphi\varphi}}{2\sqrt{g_{rr}(g^2_{t\varphi}-g_{\varphi\varphi}g_{tt})}},\\\\
    D&=\frac{(g'_{t\varphi})^2-g'_{tt}g'_{\varphi\varphi}}{4g_{rr}(g^2_{t\varphi}-g_{\varphi\varphi}g_{tt})}.
    %\Delta&=\sqrt{g_{rr}(g^2_{t\varphi}-g_{\varphi\varphi}g_{tt})}
    \end{aligned}
    \end{equation}
    Now, from the normalization condition in Eq.~(\ref{s3e7}), we obtain~\cite{Toshmatov:2019bda} 
    \begin{equation}
    \label{s3ae8}
    p^2_r=g_{rr}(-g^{tt}p^2_t-g^{\varphi\varphi}p^2_\varphi-2g^{t\varphi}p_tp_\varphi-m^2). 
    \end{equation}   
    After substituting Eqs.~(\ref{s3ae6}) and (\ref{s3e7}) into Eq.~(\ref{s3ae8}), ones obtains a second-degree polynomial equation for the energy $E$
    \begin{equation}
    \label{s3ea9}
    p^2_r=\frac{\beta}{\alpha}\left(E^2+\frac{\delta L}{\beta}E+\frac{\sigma L^2}{\beta}-\frac{\rho}{\beta}\right),
    \end{equation}
    which can be expressed as 
    \begin{equation}
    \label{s3ae10}
    p^2_r=\frac{\beta}{\alpha}(E-V_{\rm eff}^+)(E-V_{\rm eff}^-),
    \end{equation}
    where
    \begin{equation}
    \label{s3ae11}
    V_{\rm eff}^{\pm}=-\frac{\delta L}{2\beta}\pm\sqrt{\left(\frac{\delta L}{2\beta}\right)^2+\left(\frac{\rho}{\beta}-\frac{\sigma L^2}{\beta}\right)}.
    \end{equation}
    The expressions for $\alpha$, $\beta$, $\delta/2$, $\sigma$ and $\rho$ are given in Ref.~\cite{Toshmatov:2019bda}. Nevertheless, it is important to note that we found some typos in the expressions defining $\alpha$, $\delta$ and $\sigma$. In the definition of $\alpha$, it should be $g^{rr}$ instead $g_{rr}$. In the expression for $\delta$, the second term in the numerator where $s^2$ is the factor, one should have $g^{t\varphi}g'_{tt}$ instead of $g^{t\varphi}g_{tt}$. Finally, in the expression defining $\sigma$, the minus sign in the second term of the numerator, where $s$ is the factor, one should has $+g^{\varphi\varphi}g'_{t\varphi}$ instead of $-g^{\varphi\varphi}g'_{t\varphi}$. Hence, we obtain the following expressions
    \newpage
    \begin{widetext}
    \begin{equation}
    \label{s3ae12}
    \begin{aligned}
    \alpha&=g^{rr}(1-Ds^2)^2,\\\\
    \beta&=-g^{tt}+\frac{s(g^{t\varphi}g'_{\varphi\varphi}+g^{tt}g'_{t\varphi})}{\sqrt{g_{rr}(g^2_{t\varphi}-g_{\varphi\varphi}g_{tt})}}-\frac{s^2\left[g^{tt}(g'_{t\varphi})^2+g'_{\varphi\varphi}(g^{\varphi\varphi}g'_{\varphi\varphi}+2g^{t\varphi}g'_{t\varphi})\right]}{4g_{rr}(g^2_{t\varphi}-g_{\varphi\varphi}g_{tt})}, \\\\
    \frac{\delta}{2}&=g^{t\varphi}+\frac{s(g^{tt}g'_{tt}-g^{\varphi\varphi}g'_{\varphi\varphi})}{2\sqrt{g_{rr}(g^2_{t\varphi}-g_{\varphi\varphi}g_{tt})}}-\frac{s^2\left[g'_{t\varphi}(g^{tt}g'_{tt}+g^{t\varphi}g'_{t\varphi})+g'_{\varphi\varphi}(g^{t\varphi}g'_{tt}+g^{\varphi\varphi}g'_{t\varphi})\right]}{4g_{rr}(g^2_{t\varphi}-g_{\varphi\varphi}g_{tt})},\\\\
    \sigma&=-g^{\varphi\varphi}-\frac{s(g^{t\varphi}g'_{tt}+g^{\varphi\varphi}g'_{t\varphi})}{\sqrt{g_{rr}(g^2_{t\varphi}-g_{\varphi\varphi}g_{tt})}}-\frac{s^2\left[g^{tt}(g'_{tt})^2+g'_{t\varphi}(g^{\varphi\varphi}g'_{t\varphi}+2g^{t\varphi}g'_{tt})\right]}{\sqrt{g_{rr}(g^2_{t\varphi}-g_{\varphi\varphi}g_{tt})}},\\\\
    \rho&=m^2(1-Ds^2)^2.
    \end{aligned}
    \end{equation}
    \end{widetext}

    \subsection{Superluminal bound\label{SecIIIB}}

	In this section, we focus our attention on the well-known \textit{superluminal bound}. This constraint will be crucial to find the allowed values of $s$ to keep a spinning test particle moving in a trajectory with physical meaning. In this sense, we will obtain analytic expressions in the case of static and axially symmetric space-times. Then, in Sec.~\ref{SecIV}, we will use these formulas in the particular case of a traversable rotating wormhole.

	As we pointed out above, although $p_\alpha p^\alpha = -m^2$ is satisfied, the normalization $u_\alpha u^\alpha = -1$ does not necessarily hold because the four-momentum and the four-velocity are not always parallel to each other. Therefore, as the spinning test particle moves closer to the center of symmetry, the four-velocity $u^\alpha$ increases, and for specific values of spin $s$ and radius, some components of $u^\alpha$ may diverge. In this sense, the motion of the spinning particle crosses the boundary between time-like and space-like trajectories, becoming in this way superluminal. 

	Particles moving in a space-like trajectory (superluminal motion) do not have a physical meaning. Therefore, the transition to $u_\alpha u^\alpha > 0$ is not allowed for real particles. As a consequence, one must impose a further constraint: \textit{the superluminal bound}, defined by the condition $u_\alpha u^\alpha = 0$. Hence, to keep spinning test particles moving in space-like trajectories, it is necessary to impose the following constrain~\cite{Toshmatov:2019bda, Conde:2019juj} 
    \begin{equation}
    \label{s3be1}
    \frac{u_\alpha u^\alpha}{(u^t)^2}=g_{tt}+g_{rr}\left(\frac{dr}{dt}\right)^2+2g_{t\varphi}\frac{d\varphi}{dt}+g_{\varphi\varphi}\left(\frac{d\varphi}{dt}\right)^2\leq 0.
    \end{equation}
    To consider the superluminal bound in our investigation, we need to obtain analytical expressions for $dr/dt $ and $d\varphi/dt$. In order to do so, we follow a method proposed by Hojman and Asenjo in Ref.~\cite{Hojman:2012me}.

	In Sec.\ref{SecIIIA}, we constrain our calculations to the equatorial plane ($\theta=\pi/2$). Ergo, the non-vanishing components of $S^{\alpha\beta}$ are $S^{tr}$, $S^{t\varphi}$ and $S^{r\varphi}$. Hence, from the second MPD equation in Eq.~(\ref{s3e1}), we obtain the following system of equations
    \begin{equation}
    \label{s3be2}
    \begin{aligned}
    \frac{DS^{tr}}{d\lambda}&=p^tu^r-u^tp^r,\\\\\frac{DS^{t\varphi}}{d\lambda}&=p^tu^\varphi-u^tp^\varphi,\\\\
    \frac{DS^{\varphi r}}{d\lambda}&=p^\varphi u^r-u^\varphi p^r.
    \end{aligned}
    \end{equation}
    In order to solve the last system, we note that it is possible to reduce it into a system of two equation by expressing the first and second equations in terms of $S^{\varphi r}$. To do so, one needs to consider the Tulczyjew-SSC condition $S^{\alpha\beta}p_\alpha=0$. In the case of $S^{tr}$, for example, Eq.~(\ref{s3e6}) reduces to
    \begin{equation}
    \label{s3be3}
    S^{tr}p_t+S^{\varphi r}p_\varphi=0.
    \end{equation}
    Then, after applying the operator $D/d\lambda$, using the second relation in Eq.~(\ref{s3ae3}), solving for $DS^{tr}/d\lambda$, and replacing in the first equation of Eq.~(\ref{s3be2}), we obtain~\cite{Benavides-Gallego:2021lqn}
    \begin{eqnarray}
    \nonumber
    \label{s3be4}
    \frac{DS^{tr}}{d\lambda}&=&\frac{S^{\varphi r}}{p_t}\left(\frac{p_\varphi}{pt}\frac{Dp_t}{d\lambda}-\frac{Dp_\varphi}{d\lambda}\right)-\frac{p_\varphi}{p_t}\frac{DS^{\varphi r}}{d\lambda}\\ &=&p^tu^r-u^tp^r.
    \end{eqnarray}
    
    We proceed similarly with $DS^{t\varphi}/d\lambda$ in Eq.~(\ref{s3be2}), obtaining~\cite{Benavides-Gallego:2021lqn}
   
    \begin{eqnarray}
    \label{s3be5}
    \nonumber
    \frac{DS^{t\varphi}}{d\lambda}&=&\frac{p_r}{p_t}\frac{DS^{\varphi r}}{d\lambda}+\frac{S^{\varphi r}}{p_t}\left(\frac{Dp_r}{d\lambda}-\frac{p_r}{p_t}\frac{Dp_t}{d\lambda}\right)\\&=&p^tu^\varphi-u^t p^\varphi.
    \end{eqnarray}
    
    With the MPD equation for $S^{tr}$ and $S^{t\varphi}$ expressed in terms of $DS^{\varphi r}/d\lambda$, we can now use the MPD for $S^{\varphi r}$ (third equation in Eq.~(\ref{s3be2})) to reduce the system from three to only two equations. This new system is given by~\cite{Benavides-Gallego:2021lqn}
    %\begin{widetext}
    \begin{equation}
    \label{s3be6}
    \begin{aligned}
    \frac{S^{\varphi r}}{p_t}\left(p_\varphi\frac{Dp_t}{d\lambda}-p_t\frac{Dp_\varphi}{d\lambda}\right)&=
    u^r\left(p_\varphi p^\varphi + p_tp^t\right)\\
    &-p^r(p_\varphi u^\varphi+p_tu^t),\\
    \frac{S^{\varphi r}}{p_t}\left(p_t\frac{Dp_r}{d\lambda}-p_r\frac{Dp_t}{d\lambda}\right)&=u^\varphi\left(p_rp^r+p_tp^t\right)\\
    &-p^\varphi\left(p_ru^r+p_tu^t\right).
    \end{aligned}
    \end{equation}
    %\end{widetext}
    Now, using the first MPD equation in Eq.~(\ref{s3be2}), we obtain the following relations (see Appendix~\ref{appenB})
    \begin{equation}
    \label{s3be7}
    \begin{aligned}
    \frac{Dp_t}{d\lambda}&=\frac{S^{\varphi r}}{p_t}\left[\left(p_\varphi R_{trtr}-p_t R_{tr\varphi r}\right)u^r-p_r R_{t\varphi t\varphi}u^\varphi\right],\\
    \frac{Dp_r}{d\lambda}&=\left[(p_\varphi R_{rttr}+p_tR_{rtr\varphi})u^t+(p_\varphi R_{r\varphi tr}+p_t R_{r\varphi r\varphi})u^\varphi\right]\\
    &\times\frac{S^{\varphi r}}{p_t}\\\\ 
    %\frac{S^{\varphi r}}{p_t}\left[p_\varphi R_{rttr}u^t+(p_\varphi R_{r\varphi tr}+p_t R_{r\varphi r\varphi})u^\varphi\right],\\\\
    %
    \frac{Dp_\varphi}{d\lambda}&=\frac{S^{\varphi r}}{p_t}
    \left[u^r(p_\varphi R_{\varphi rtr}+p_t R_{\varphi rr\varphi})-p_r R_{\varphi tt\varphi}u^t\right].
    \end{aligned}
    \end{equation}
    After replacing Eq.~(\ref{s3be7}) into Eq.~(\ref{s3be6}), we obtain the following system
    \begin{equation}
    \label{s3be8}
    \begin{aligned}
    u^r\left[p^2_\varphi\mathcal{A}+2p_\varphi p_t \mathcal{D}+p^2_t \mathcal{B}\right]=&u^\varphi \mathcal{C}p_\varphi p_r + u^t\mathcal{C}p_tp_r,\\\\
    u^\varphi\left[p^2_t\mathcal{B}+p_t p_\varphi\mathcal{D}+p^2_r\mathcal{C}\right]=&u^r\left[\mathcal{D}p_r p_t+\mathcal{A}p_r p_\varphi\right]+\\
    &u^t\left[\mathcal{A}p_tp_\varphi+p^2_t \mathcal{D}\right],
    \end{aligned}
    \end{equation}
    where 
    \begin{equation}
    \label{s3be9}
    \begin{aligned}
    \mathcal{\hat{A}}&=g^{\varphi\varphi}+\left(\frac{S^{\varphi r}}{p_t}\right)^2R_{trrt},\\\\
    \mathcal{\hat{B}}&=g^{tt}+\left(\frac{S^{\varphi r}}{p_t}\right)^2R_{\varphi rr \varphi},\\\\
    \mathcal{\hat{C}}&=g^{rr}+\left(\frac{S^{\varphi r}}{p_t}\right)^2R_{\varphi tt \varphi}.\\\\
    \mathcal{\hat{D}}&=g^{t\varphi}+\left(\frac{S^{\varphi r}}{p_t}\right)^2R_{tr\varphi r}.
    \end{aligned}
    \end{equation}
    Hence, after following the gauge choices and invariant relations in Ref~\cite{Hojman:2012me}, we can solve the above system of equations to obtain
    %\begin{widetext}
    \begin{equation}
    \label{s3be10}
    \begin{aligned}
    \frac{dr}{dt}=\frac{u^r}{u^t}&=\frac{\mathcal{C}p_r}{\mathcal{B}p_t +\mathcal{D}p_\varphi},\\\\
    %\frac{\mathcal{C} p_r \left\{p_\varphi \left[\mathcal{A} p_\varphi+p_t (\mathcal{D}+g^{t\varphi})\right]+\mathcal{B} p^2_t+\mathcal{C} p^2_r\right\}}{\left[\mathcal{B} p_t+\mathcal{D} p_\varphi\right] \left[p_\varphi \mathcal{K}+B p^2_t+\mathcal{C} p^2_r\right]},\\
     % 
    \frac{d\varphi}{dt}=\frac{u^\varphi}{u^t}&=\frac{\mathcal{D}p_t+\mathcal{A}p_\varphi}{\mathcal{B}p_t +\mathcal{D}p_\varphi}.
    %\frac{\mathcal{C}p^2_r\left(\mathcal{D}p_t+\mathcal{A}p_\varphi\right)+\left(g^{t\varphi}p_t+\mathcal{A}p_\varphi\right)\left[\mathcal{B}p^2_t+p_\varphi\mathcal{K}\right]}{\left(\mathcal{B}p_t+\mathcal{D} p_\varphi\right)\left[\mathcal{C}p^2_r + \mathcal{B} p^2_t+p_\varphi\mathcal{K}\right]},\\
    %
    %\mathcal{K}&=\mathcal{A} p_\varphi+2\mathcal{D}p_t.
    \end{aligned}
    \end{equation}
    %\end{widetext}
    Note that Eqs.~(\ref{s3be9}) and (\ref{s3be10}) reduces to Eqs.~(42) and (41) of Ref.~\cite{Benavides-Gallego:2021lqn} when $R_{tr\varphi r}=0$ and $g^{t\varphi}=0$, e.i. when $\mathcal{D}=0$. 
    
%%%%%%%%%%%%%%%%%%%%%%%%%%%%%%%%%%%%%%%%%%%%%%%%%%%%%%%%%%% %%%%%%%%%%%%%% Section IV %%%%%%%%%%%%%%%%%%%%%%%%%%%%%%%%%
%%%%%%%%%%%%%%%%%%%%%%%%%%%%%%%%%%%%%%%%%%%%%%%%%%%%%%%%%%%
    
    \section{Dynamics of spinning test particles around a rotating wormhole\label{SecIV}} 
    
    The results of Sec.\ref{SecII} and Sec.\ref{SecIII} describe the dynamics of a spinning test particle in a static and axially symmetric space-time. In this section, we apply these results to the geometry of a rotating wormhole. 

    \subsection{Effective potential}
    
    We begin by considering the canonical form of Eq.~(\ref{s2be1}) with the following functions 
    \begin{equation}
    \label{s4ae1}
    \begin{array}{cccc}
    N&=e^{\Phi},&\Phi=-\frac{b_0}{r},&\\\\
    K&=1,&\omega=\frac{2J}{r^3},&\\\\
    b&=\frac{b^2_0}{r},&\mu=-\ln\left(1-\frac{b}{r}\right).&
    \end{array}
    \end{equation}
    Hence, the space-time takes the form
    \begin{equation}
    \label{s4ae2}
    \begin{aligned}
    ds^2=&-\left(e^{2\Phi}-\omega^2r^2 \sin^2\theta \right)dt^2-2\omega r^2\sin^2\theta dt d\varphi\\
    &+\left(1-\frac{b^2_0}{r^2}\right)^{-1}dr^2+r^2(d\theta^2+\sin^2\theta d\varphi^2).
    \end{aligned}
    \end{equation}
    Here, $b_0$ is the wormhole throat, which is interpreted as the wormhole mass. When $\omega=0$, note that the space-time reduces to the Morris-Thorne space-time~\cite{Morris:1988cz}. Using the coordinate transformation $r^2=l^2+b^2_0$, Eq.~(\ref{s4ae2}) reduces to
    \begin{equation}
    \label{s4ae3}
    \begin{aligned}
    ds^2=&-\left(e^{2\Phi}-\omega^2(l^2+b^2_0) \sin^2\theta\right)dt^2\\
    &-2\omega (l^2+b^2_0)\sin^2\theta dt d\varphi\\
    &+dl^2+(l^2+b^2_0)(d\theta^2+\sin^2\theta d\varphi^2),
    \end{aligned}
    \end{equation}
    with $\Phi$ and $\omega$ now functions of $l$. In this new radial coordinate, the throat of the wormhole is at $l=0$ ($r_0=b_0$). 

	Before computing the effective potential, we want to express the results in terms of dimensionless variables, where  
    \begin{equation}
    \label{s4ae4}
    \begin{array}{cccc}
    l\rightarrow\frac{l}{b_0},&s\rightarrow\frac{s}{b_0}=\frac{S}{m b_0},&\mathcal{L}\rightarrow\frac{\mathcal{L}}{b_0}=\frac{L}{m b_0},&a\rightarrow \frac{J}{b^2_0}.
    \end{array}
    \end{equation}
    Therefore, the effective potential has the following form
    \begin{equation}
    \label{s4ae4a}
    \mathcal{V}_{\rm eff}=-\frac{\delta \mathcal{L}}{2\beta}\pm\sqrt{\left(\frac{\delta \mathcal{L}}{2\beta}\right)^2+\left(\frac{\rho}{\beta}-\frac{\sigma \mathcal{L}^2}{\beta}\right)}.
    \end{equation}
    with $\mathcal{V}_{\rm eff}=V_{\rm eff}/m$\footnote{From now on, we use $\mathcal{V}_\text{eff}$ with $+$ in Eq.~(\ref{s4ae4a}).}. Then, after using the definitions in Eq.~(\ref{s3ae12}), we obtain 
    \begin{widetext}
    \begin{equation}
    \label{s4ae5}
    \begin{aligned}
    \beta&=e^{-2 \Phi}-\frac{6 a l  e^{-3 \Phi}}{\left(l^2+1\right)^2}s+\frac{l^2 e^{-4 \Phi} \left(9 a^2-\left(l^2+1\right)^2 e^{2 \Phi}\right)}{\left(l^2+1\right)^4}s^2,\\\\
    %%%%%%
    \delta\rightarrow b_0\delta&=-\frac{4 a e^{-2 \Phi}}{\left(l^2+1\right)^{3/2}}+\frac{2e^{3\Phi}\left(12a^2 l+\left(l^2+1\right)^2 e^{2 \Phi} \left(\left(l^2+1\right) \Phi'-l\right)\right)}{(l^2+1)^{7/2}}s\\
    &+\frac{2a l e^{-4 \Phi (l)} \left(-18 a^2 l-\left(l^2+1\right)^2 e^{2 \Phi} \left(3 \left(l^2+1\right) \Phi'+l\right)\right)}{\left(l^2+1\right)^{11/2}}s^2,\\\\
    %%%%%%
    \sigma\rightarrow b^2_0\sigma&=\frac{4 a^2 e^{-2 \Phi}}{\left(l^2+1\right)^3}-\frac{1}{l^2+1}-\frac{2ae^{-3\Phi}\left(12 a^2 l+\left(l^2+1\right)^2 e^{2 \Phi} \left(2 \left(l^2+1\right) \Phi'+l\right)\right)}{(l^2+1)^5}s\\
    &+\frac{a^2 l e^{-4 \Phi } \left(36 a^2 l+\left(l^2+1\right)^2 e^{2 \Phi } \left(12 \left(l^2+1\right) \Phi'-l\right)\right)+\left(l^2+1\right)^6 \Phi'^2}{(l^2+1)^7}s^2,\\\\
    %%%%%%
    \rho\rightarrow\frac{\rho}{m^2}&=s^2 \left(-\frac{18 a^2 l^2 e^{-2 \Phi}}{\left(l^2+1\right)^4}-\frac{2 l \Phi'}{l^2+1}\right)+s^4 \left(\frac{81 a^4 l^4 e^{-4 \Phi}}{\left(l^2+1\right)^8}+\frac{18 a^2 l^3 e^{-2 \Phi} \Phi'}{\left(l^2+1\right)^5}+\frac{l^2 \Phi'^2}{\left(l^2+1\right)^2}\right)+1.
    \end{aligned}
    \end{equation}
    \end{widetext}
    It is straightforward to check that the last expressions reduce to those in Ref.~\cite{Benavides-Gallego:2021lqn} when $a=0$.

\begin{figure}[t]
    \begin{center}
    \includegraphics[scale=0.4]{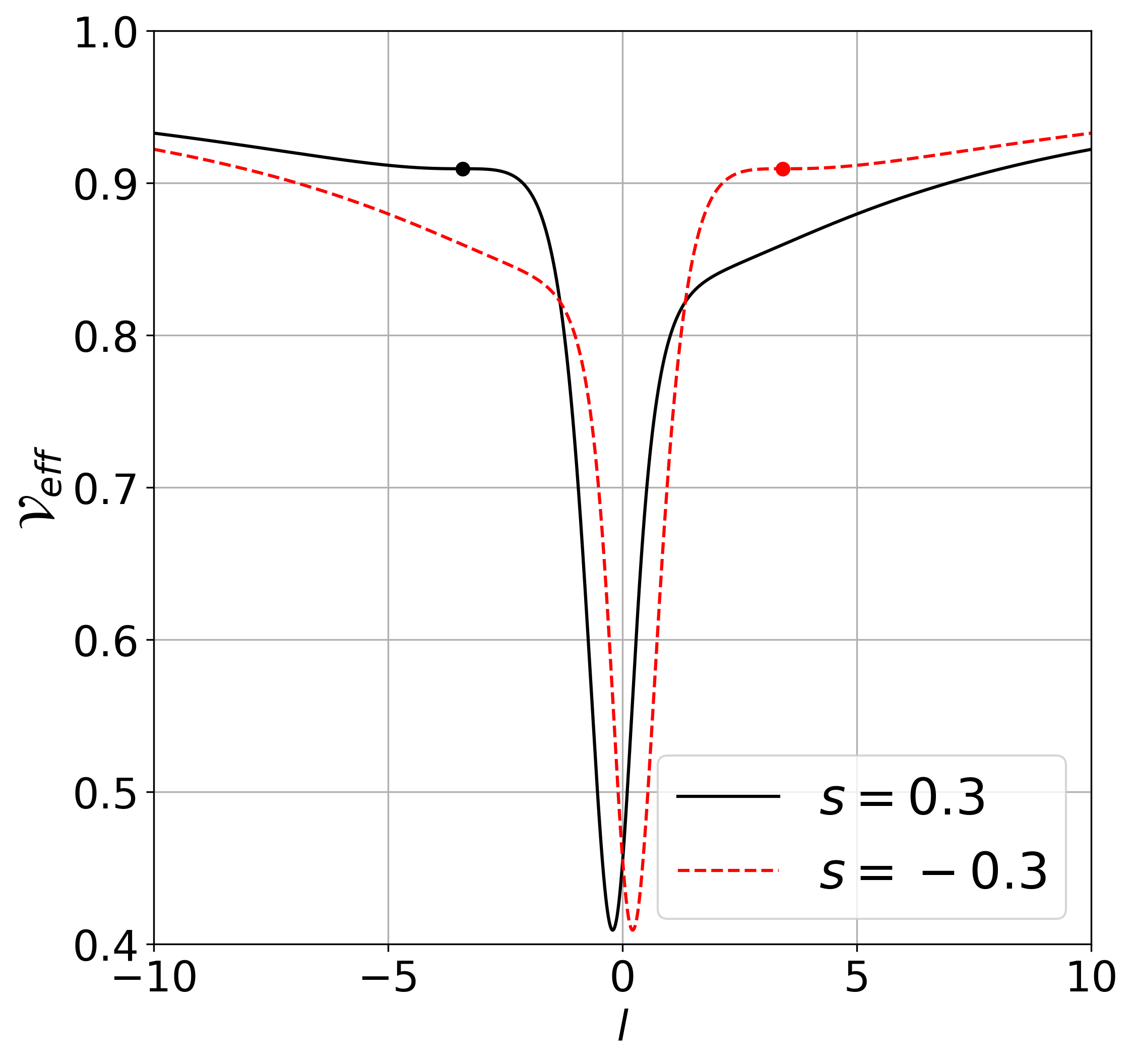}
    \caption{The ``\textit{plus}'' (black) and ``\textit{minus}'' (red) configurations for the effective potential. The ISCO for each configuration is shown using dots. We assume $b_0=M=1$.\label{fig1}}
    \end{center}
    \end{figure}

	From Eqs.~(\ref{s4ae4}) and (\ref{s4ae5}), it is possible to see the symmetries in $\mathcal{V}_{\text{eff}}$ depending on the signs of $s$, $a$ and $\mathcal{L}$. We call $\mathcal{V}^P_{\text{eff}}$ the ``\textit{plus}'' configuration\footnote{Since $\mathcal{V}_{\text{eff}}$ depends on the wormhole's spin $a$, we also found the following relations:
    \begin{equation*}
    \label{s5e1}
    \begin{aligned}
    \mathcal{V}^M_{\text{eff}}(l,-s,a,\mathcal{L})&=\mathcal{V}_{\text{eff}}(l,s,-a,-\mathcal{L})=\mathcal{V}_{\text{eff}}(-l,s,a,\mathcal{L})\\
    &=\mathcal{V}_{\text{eff}}(-l,-s,-a,-\mathcal{L}),
    \end{aligned}
    \end{equation*}
    and 
    \begin{equation*}
    \label{s5e2}
    \begin{aligned}
    \mathcal{V}^P_{\text{eff}}(l,s,a,\mathcal{L})&=\mathcal{V}_{\text{eff}}(l,-s,-a,-\mathcal{L})=\mathcal{V}_{\text{eff}}(-l,-s,a,\mathcal{L})\\
    &=\mathcal{V}_{\text{eff}}(-l,s,-a,-\mathcal{L}).
    \end{aligned}
    \end{equation*}
    }
    \begin{equation}
    \label{s4ae6}
    \begin{aligned}
    \mathcal{V}^P_{\text{eff}}(l,s,a,\mathcal{L})&=\mathcal{V}^P_{\text{eff}}(l,-s,-a,-\mathcal{L}),
    \end{aligned}   
    \end{equation}
    and $\mathcal{V}^M_{\text{eff}}$ the ``minus'' configuration 
    \begin{equation}
    \label{s4ae7}
    \begin{aligned}
    \mathcal{V}^M_{\text{eff}}(l,-s,a,\mathcal{L})&=\mathcal{V}^M_{\text{eff}}(l,s,-a,-\mathcal{L}).
    \end{aligned}   
    \end{equation}
    The behavior of each configuration is shown in Fig.~\ref{fig1}, where we plot together $\mathcal{V}^P_{\text{eff}}$ (black) and $\mathcal{V}^M_{\text{eff}}$ (red), as functions of $l$ with $|a|=0.1$ and $|\mathcal{L}|=2.24359426)$. Note that the change from ``\textit{plus}'' to ``\textit{minus}'' configurations also changes the location of the innermost stable circular orbit (ISCO), shown in the figure with black and red dots. According to Eqs.~(\ref{s4ae6}) and (\ref{s4ae7}), we obtain the following relations
    \begin{equation}
    \label{s4ae8}
    \begin{aligned}
    \mathcal{V}^P_{\text{eff}}(l,+0.3,0.1,-2.24359)&=\mathcal{V}^P_{\text{eff}}(l,-0.3,-0.1,2.24359),\\
    \mathcal{V}^M_{\text{eff}}(l,-0.3,0.1,-2.24359)&=\mathcal{V}^M_{\text{eff}}(l,+0.3,-0.1,2.24359).
    \end{aligned}
    \end{equation} 
    %%%%%%%%%%%%%%%%%%%%
    %%%%%%%%%%%%%%%%%%%%
    %%%%%%%%%%%%%%%%%%%%
    \begin{figure*}[t]
    \begin{center}
    \includegraphics[scale=0.22]{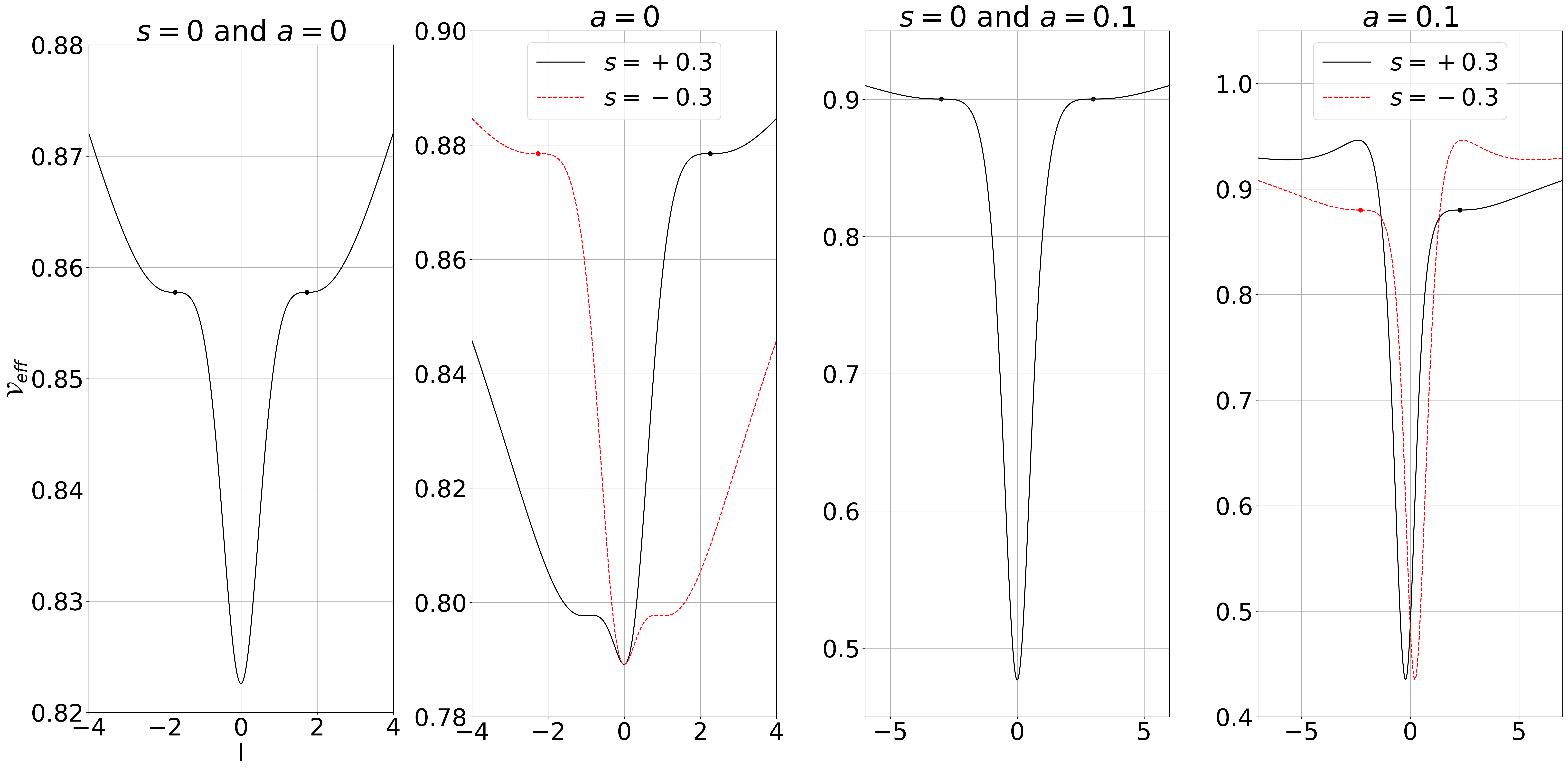}
    \caption{The effective potential in different situations. The ``\textit{plus}'' and ``\textit{minus}'' configurations are shown in black and red colors, respectively. The values for the ISCO in each configuration are shown using dots. We assume $b_0=M=1$ and $\mathcal{L}=2.0$, $1.89788245$, $-2.40282807$ and $-2.46920218$ for the first, second, third, and fourth panels respectively.\label{fig2}}
    \end{center}
    \end{figure*}
    %%%%%%%%%%%%%%%%%%%%

	Hence, from the physical point of view, in the first configuration (``\textit{plus}''), we have a system in which the particle's spin $s$ and the wormhole's spin $a$ align with the symmetry axis of the space-time, while the particle's angular momentum $\mathcal{L}$ is antiparallel. The ``\textit{minus}'' configuration, on the other hand, corresponds to the case where the particle's spin $s$ and its angular momentum $\mathcal{L}$ are parallel to the symmetry axis of the space-time, while antiparallel to the wormhole's angular momentum $\mathcal{L}$. In other words, in the ``\textit{plus}'' configuration, the particle's spin and the wormhole's spin are always antiparallel to the particle's angular momentum $\mathcal{L}$, in contrast to the ``\textit{minus}'' configuration, where the particle's spin $s$ and its angular moment $\mathcal{L}$ are always antiparallel to the wormhole's spin $a$.
	
	In Fig.\ref{fig2}, we show the behavior of the effective potential in different situations as a function of $l$. In the first panel, we show the behavior in the simplest case, i.e. when $s=a=0$. In this situation, the effective potential is symmetric, and the ISCO radius locates at the same distance from the throat for both the lower and upper universes ($l_{ISCO}=\pm1.73205081$). In the second panel, we show the shape of $\mathcal{V}_\text{eff}$ when $s=0.3$ and $a=0.0$. In this case, as showed by Benavides-Gallego et al. in Ref.~\cite{Benavides-Gallego:2021lqn}, the effective potential is not symmetric and has two possible configurations: ``\textit{plus}'' and ``\textit{minus}'', depending on the sign of the particle's spin $s$. In the ``\textit{plus}'' configuration, the ISCO situates in the upper universe. When we change to the ``\textit{minus}'' configuration, the ISCO changes from $l_{ISCO}=+2.25990751$ to $l_{ISCO}=-2.25990751$. 
	
\begin{figure}[h!]
    \begin{center}
    \includegraphics[scale=0.4]{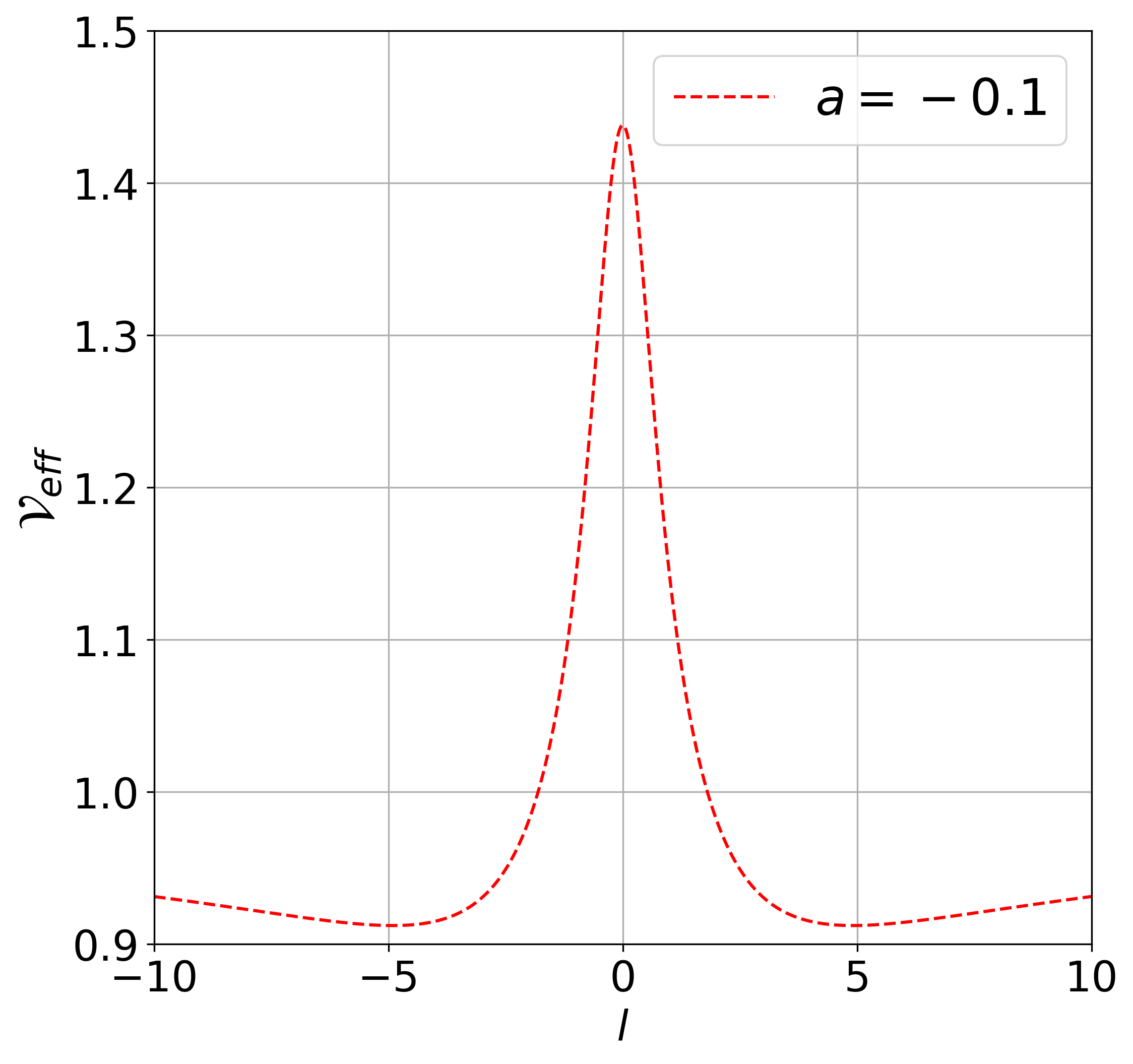}
    \caption{The ``\textit{minus}'' configuration for the third panel in Fig.~\ref{fig2}. We assume $b_0=M=1$ and $\mathcal{L}=-2.40282807$.\label{fig3}}
    \end{center}
    \end{figure}

	In the third panel, we consider the case in which $s=0.0$ and $a=0.1$. From the figure, it is possible to see that $\mathcal{V}_\text{eff}$ is once again symmetric, with the ISCO located at the same distance from the wormhole's throat in both universes ($l_{ISCO}=\pm2.9936153$). Also, note that the effective potential has a minimum value at the throat. In this particular case, changing the wormhole spin from $a=0.1$  to $a=-0.1$, the effective potential changes its shape drastically, see Fig.~\ref{fig3}. There is no ISCO in this situation.
    %%%%%%%%%%%%%%%%%%%%
    %%%%%%%%%%%%%%%%%%%%
    %%%%%%%%%%%%%%%%%%%%
    \begin{figure*}[t]
    \begin{center}
    \includegraphics[scale=0.21]{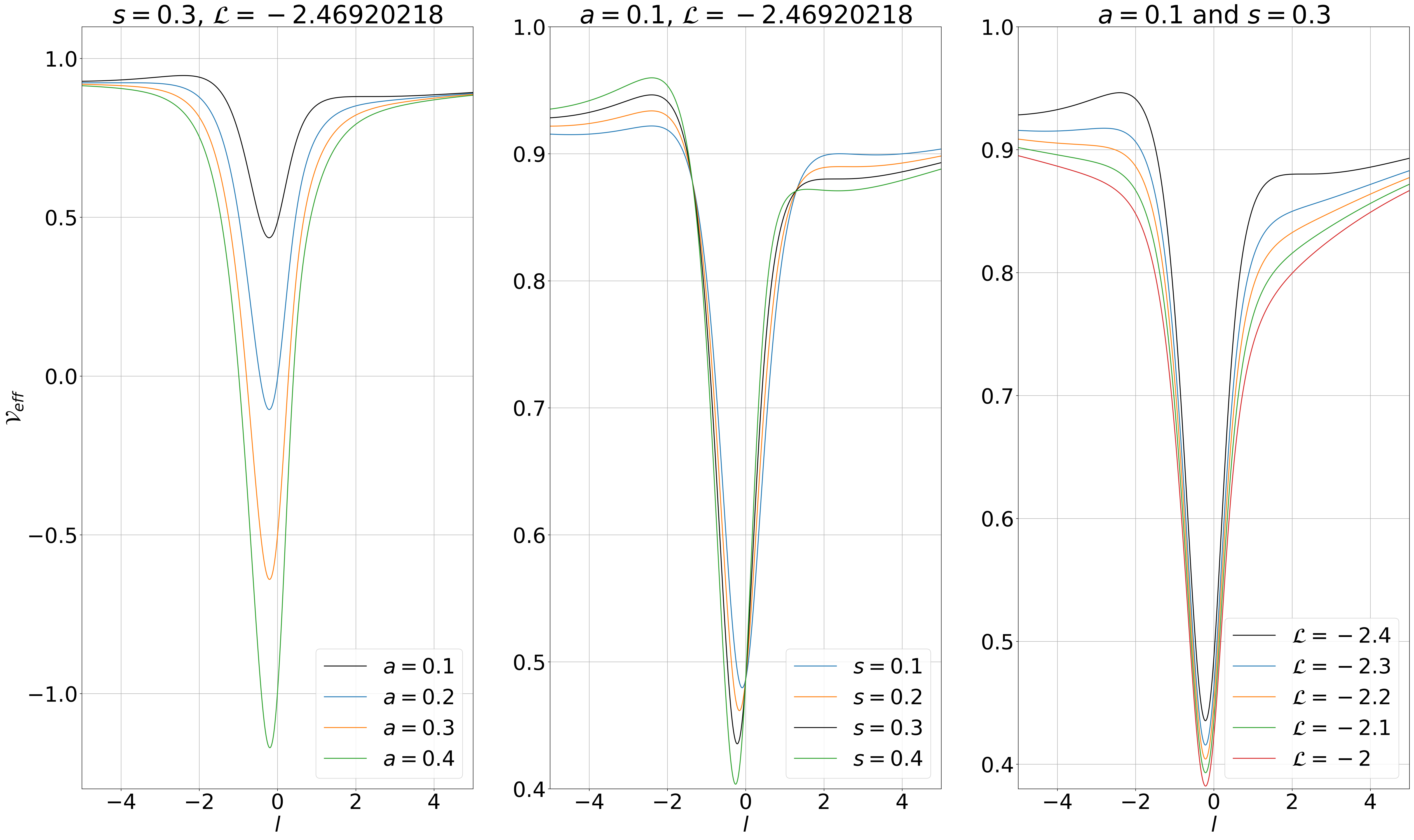}
    \caption{The effective potential in different situations. Left panel: $\mathcal{V}_{eff}$ vs. $l$ for different values of $a$. Center: $\mathcal{V}_{eff}$ vs. $l$ for different values of $s$. Right panel: $\mathcal{V}_{eff}$ vs. $l$ for different values of $\mathcal{L}$. In black color we show the case $s=0.3$, $a=0.1$ and $\mathcal{L}=-2.46920218$. We assume $b_0=M=1$.\label{fig4}}
    \end{center}
    \end{figure*}
    %%%%%%%%%%%%%%%%%%%%
    
	In the fourth panel, we show the behavior of $\mathcal{V}_\text{eff}$ in the most general case, i.e. a spinning test particle moving around a rotating wormhole. The figure shows a similar behavior as the case investigated in Ref.~\cite{Benavides-Gallego:2021lqn}: the effective potential also has two configurations (``\textit{plus}'' and ``\textit{minus}''), which change the position of the ISCO. Nevertheless, it is possible to see a difference in the shape of $\mathcal{V}_{\text{eff}}$. According to the figure, the minimum value of the effective potential shifts to the lower or upper universe depending on the configuration. In contrast to the case shown in the second panel of Fig.~\ref{fig2}, where the effective potential has a minimum value at the throat. For example, in the case shown in the fourth panel of Fig.~\ref{fig2}, $\mathcal{V}_{\text{min}}$ is shifted from the lower universe to the upper universe when we change the configuration from ``\textit{plus}'' to ``\textit{minus}''. This effect is mainly due to the wormhole's spin $a$.

    %%%%%%%%%%%%%%%%%%%%
    \begin{figure*}[t]
    \begin{center}$
    \begin{array}{ccc}
    \includegraphics[scale=0.28]{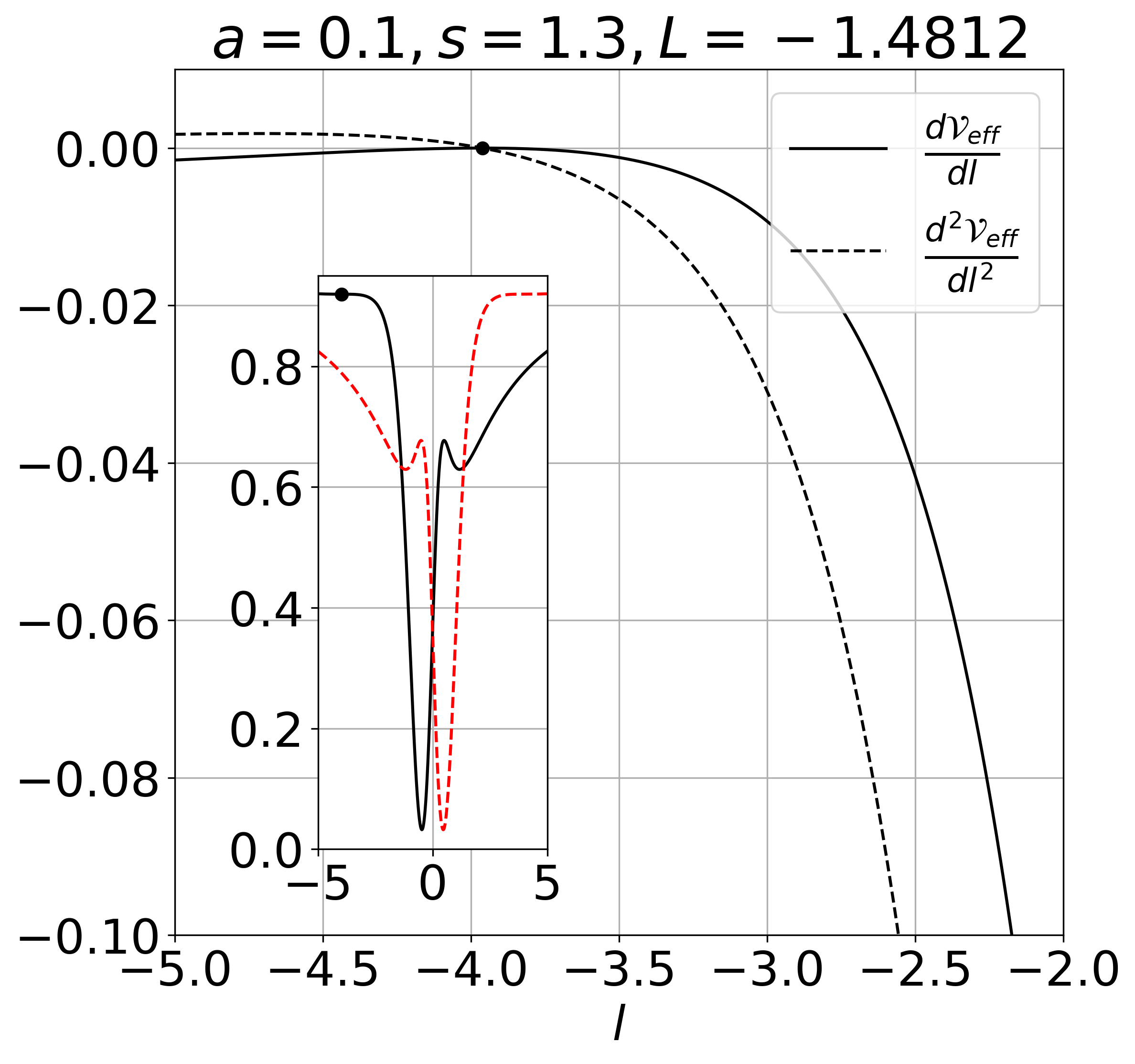}&
    \includegraphics[scale=0.28]{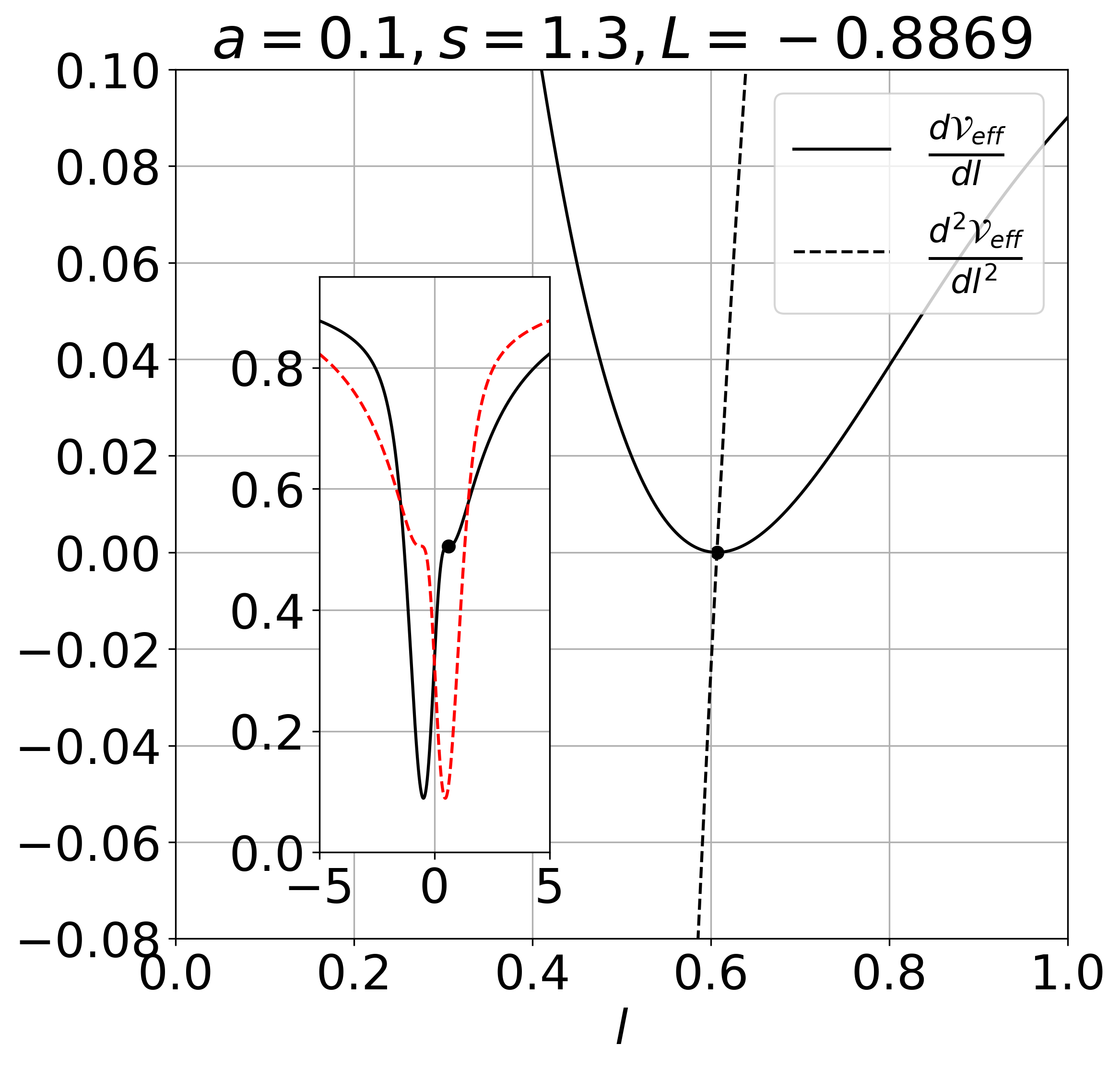}&
    \includegraphics[scale=0.28]{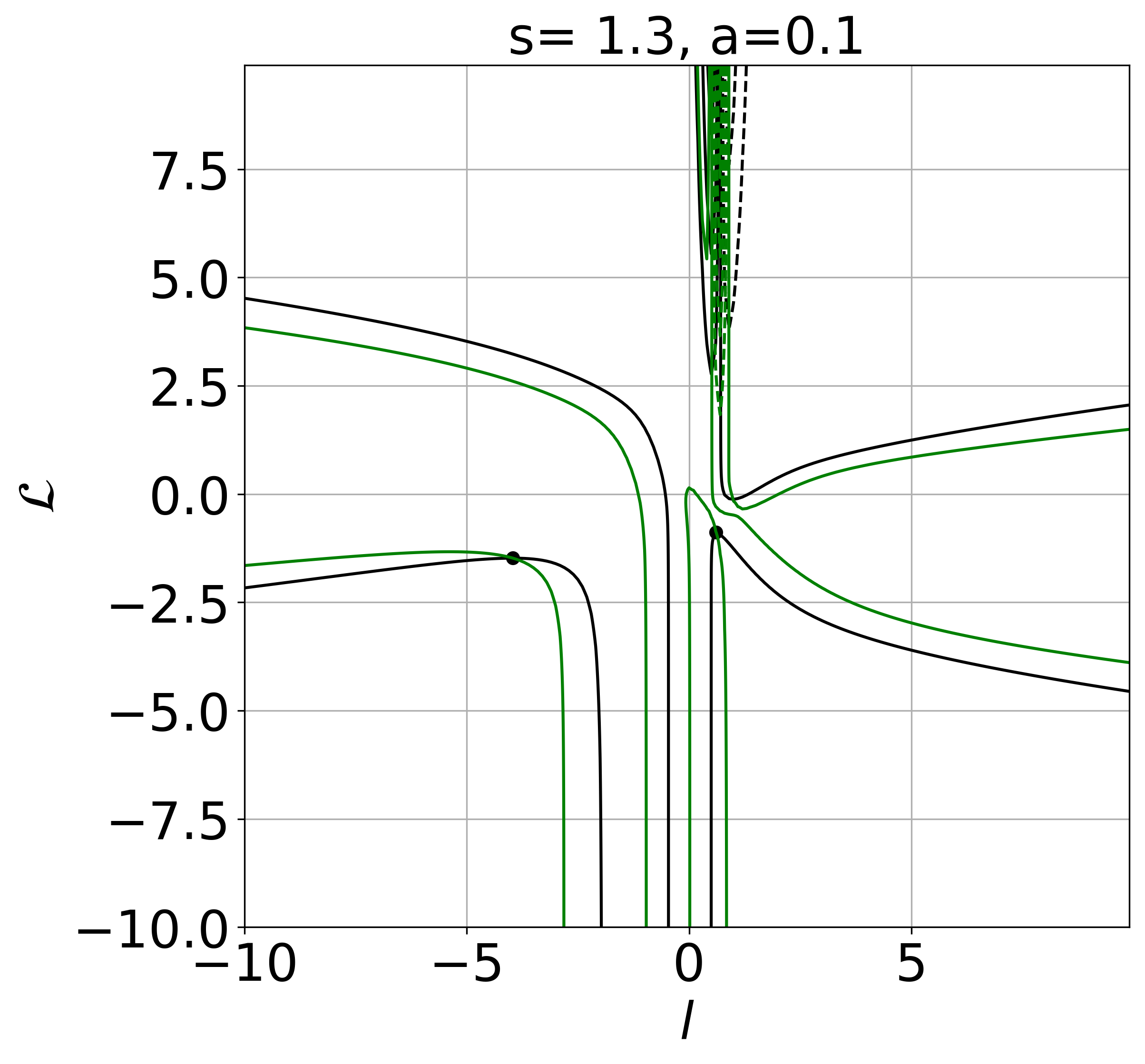}
    \end{array}$
    \end{center}
    \caption{The first (continuous black line) and second derivatives (dashed black line) of $\mathcal{V}_\text{eff}$ as a function of $l$. The intersection between the plots corresponds to $l_{ISCO}$, see the black dot in each panel. Left panel: $l_{ISCO}$ in the lower universe.  Center: $l_{ISCO}$ in the upper universe.  Right panel: Contour plot of the first (black) and second (green) derivatives of $\mathcal{V}_\text{eff}$. $l_{ISCO}$ is shown using black dots. The shape of $\mathcal{V}_\text{eff}$ vs. $l$ is shown in the small panel for the first and second panels in the figure. We assume $b_0=M=1$. \label{fig6}}
    \end{figure*}
    %%%%%%%%%%%%%%%%%%%%
    	
    %%%%%%%%%%%%%%%%%%%%
    \begin{figure*}[t]
    \begin{center}$
    \begin{array}{cc}
    \includegraphics[scale=0.55]{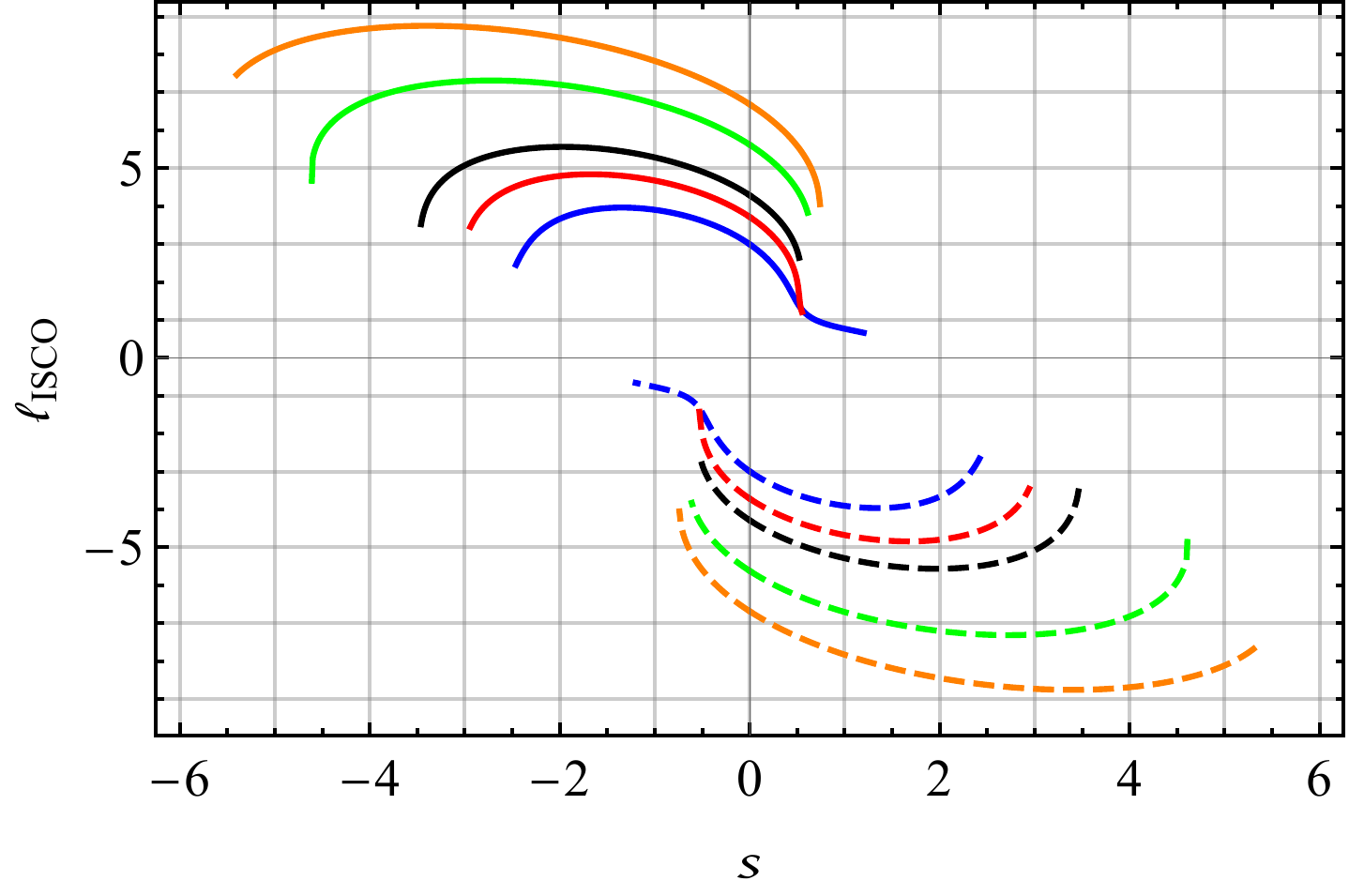}&\includegraphics[scale=0.55]{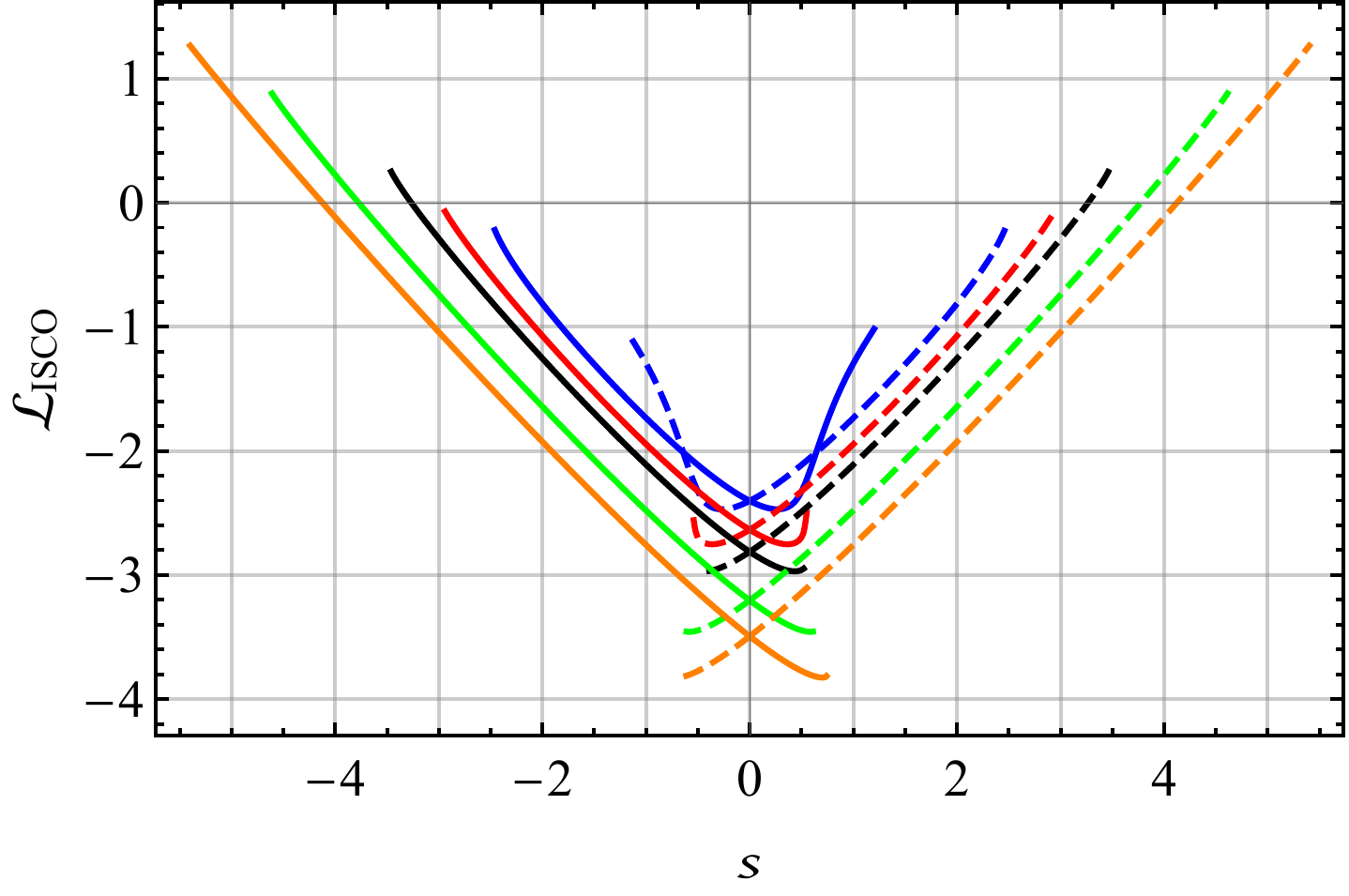}\\
    \includegraphics[scale=0.55]{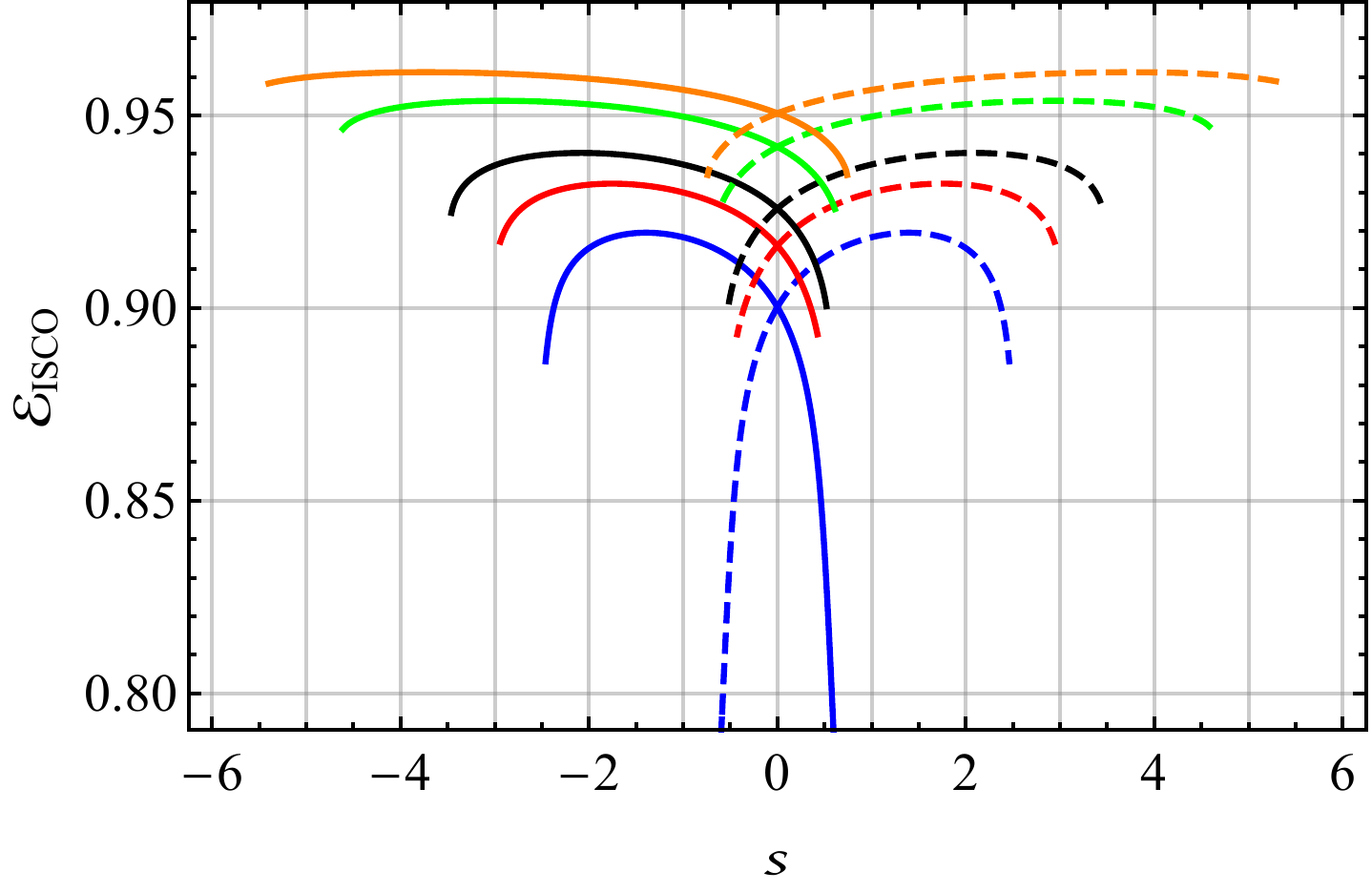}&\includegraphics[scale=0.55]{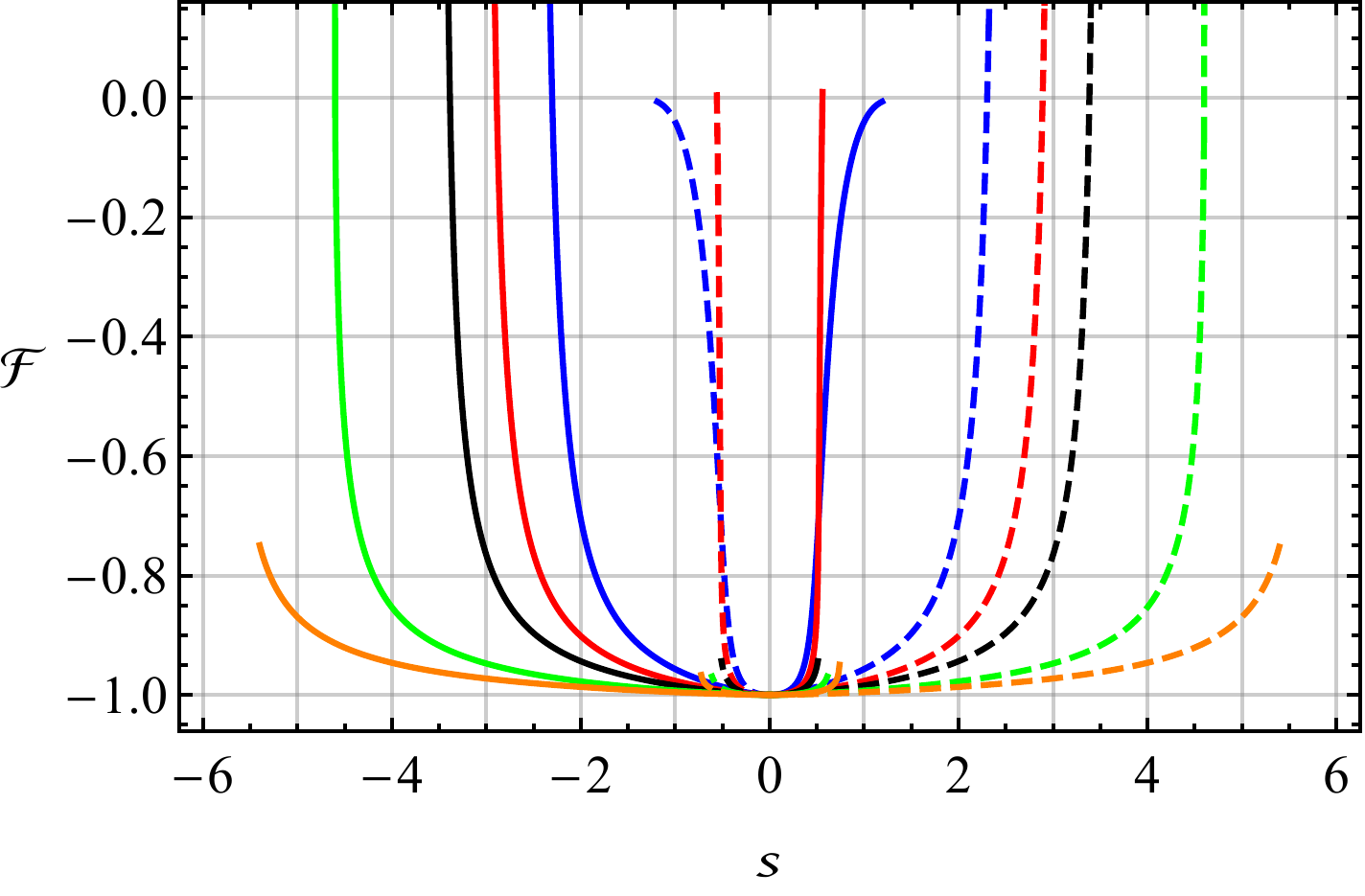}
    \end{array}$
    \end{center}
    \caption{$l_{ISCO}$, $\mathcal{L}_{ISCO}$, $\mathcal{E}_{ISCO}$ and the function $\mathcal{F}$ vs. $s$ for different values of $a=0.1$ (blue), $a=0.2$ (red), $a=0.3$ (black), $a= 0.6$ (green) and $a=0.9$ (orange) in the upper (continuous line) and lower (dashed line) universes. We assume $b_0=M=1$. \label{fig7}}
    \end{figure*}
    %%%%%%%%%%%%%%%%%%%%
    
    %%%%%%%%%%%%%%%%%%%%
    \begin{figure*}[t]
    \begin{center}$
    \begin{array}{cc}
    \includegraphics[scale=0.55]{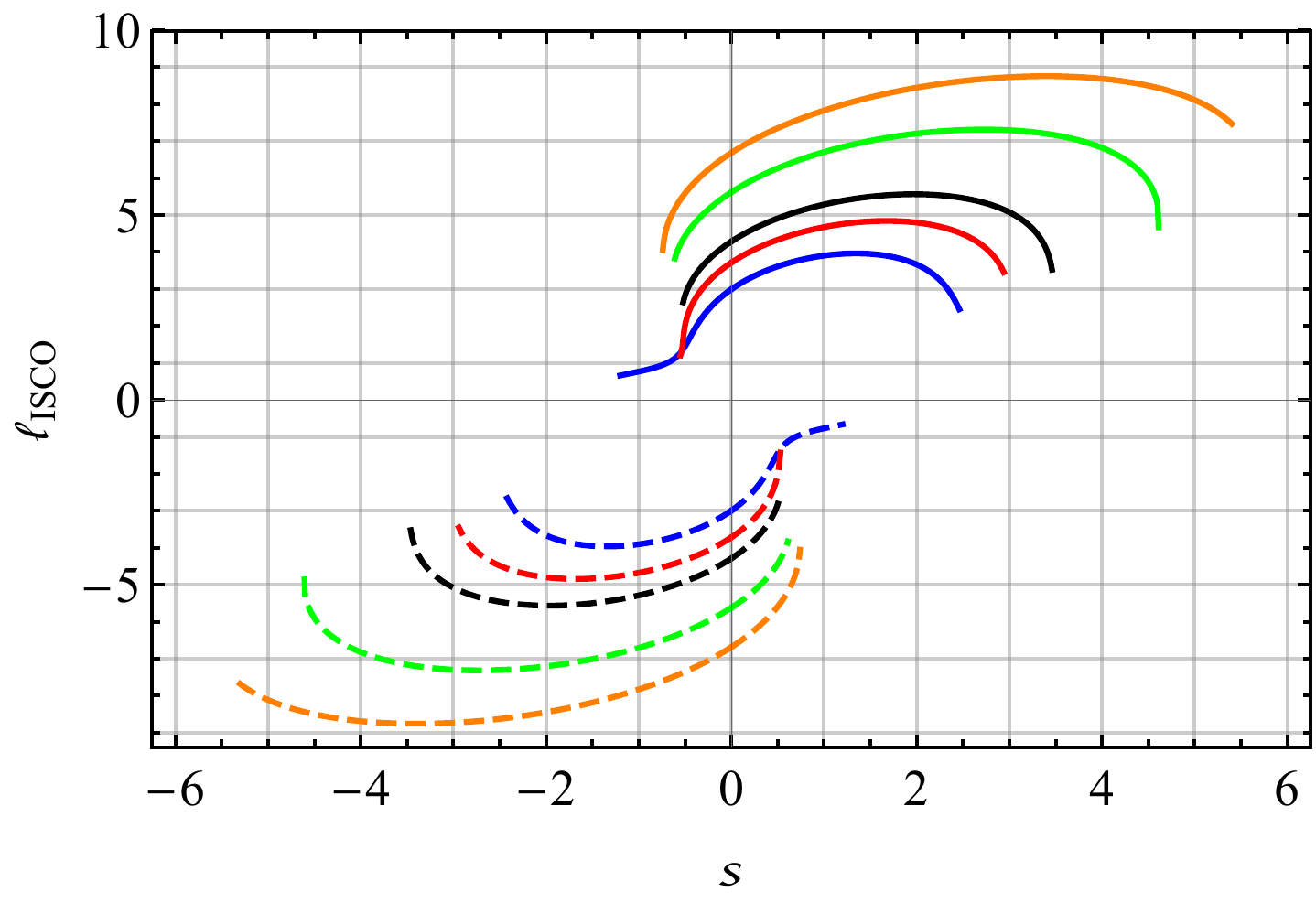}&\includegraphics[scale=0.55]{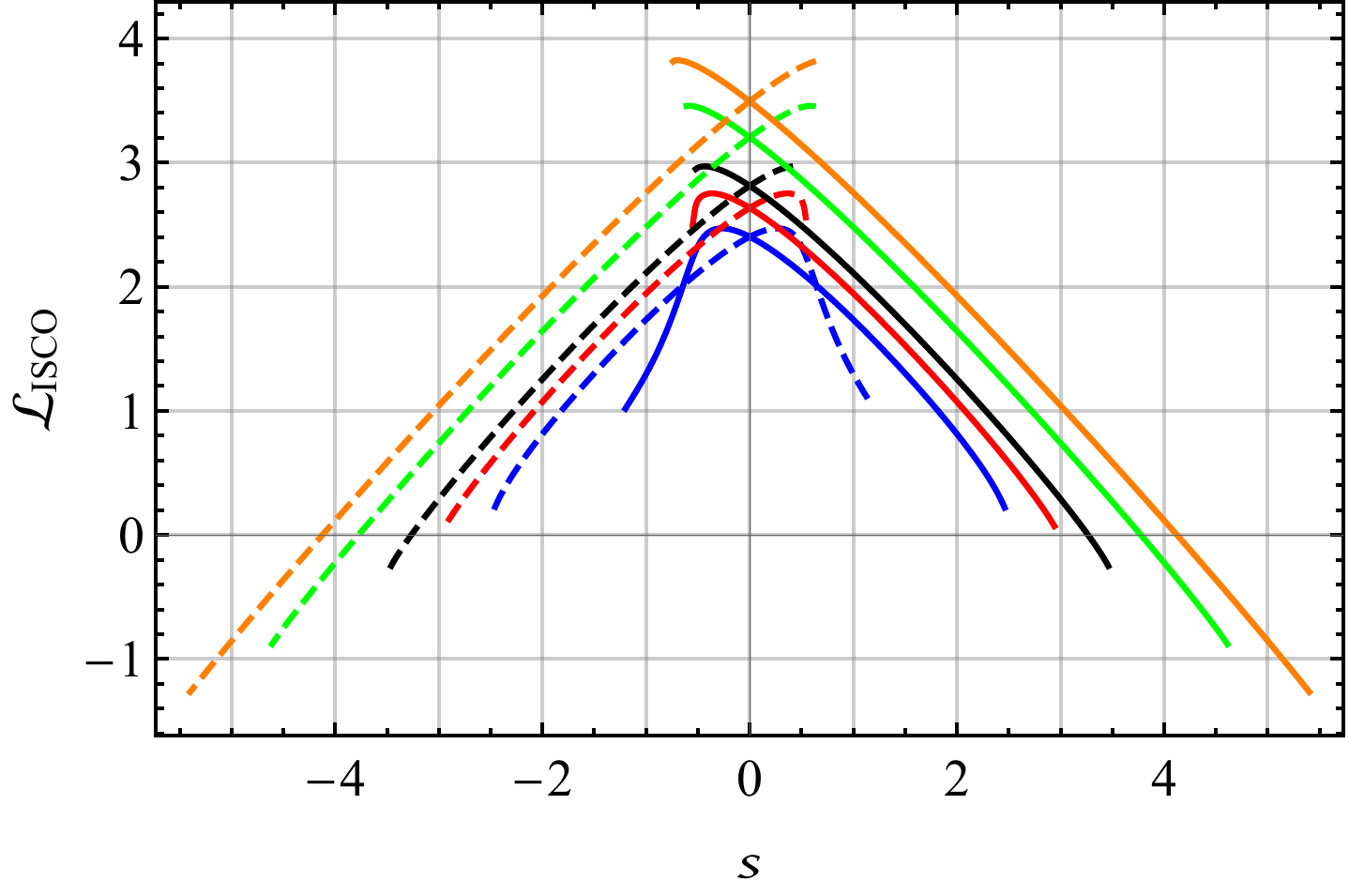}\\
    \includegraphics[scale=0.55]{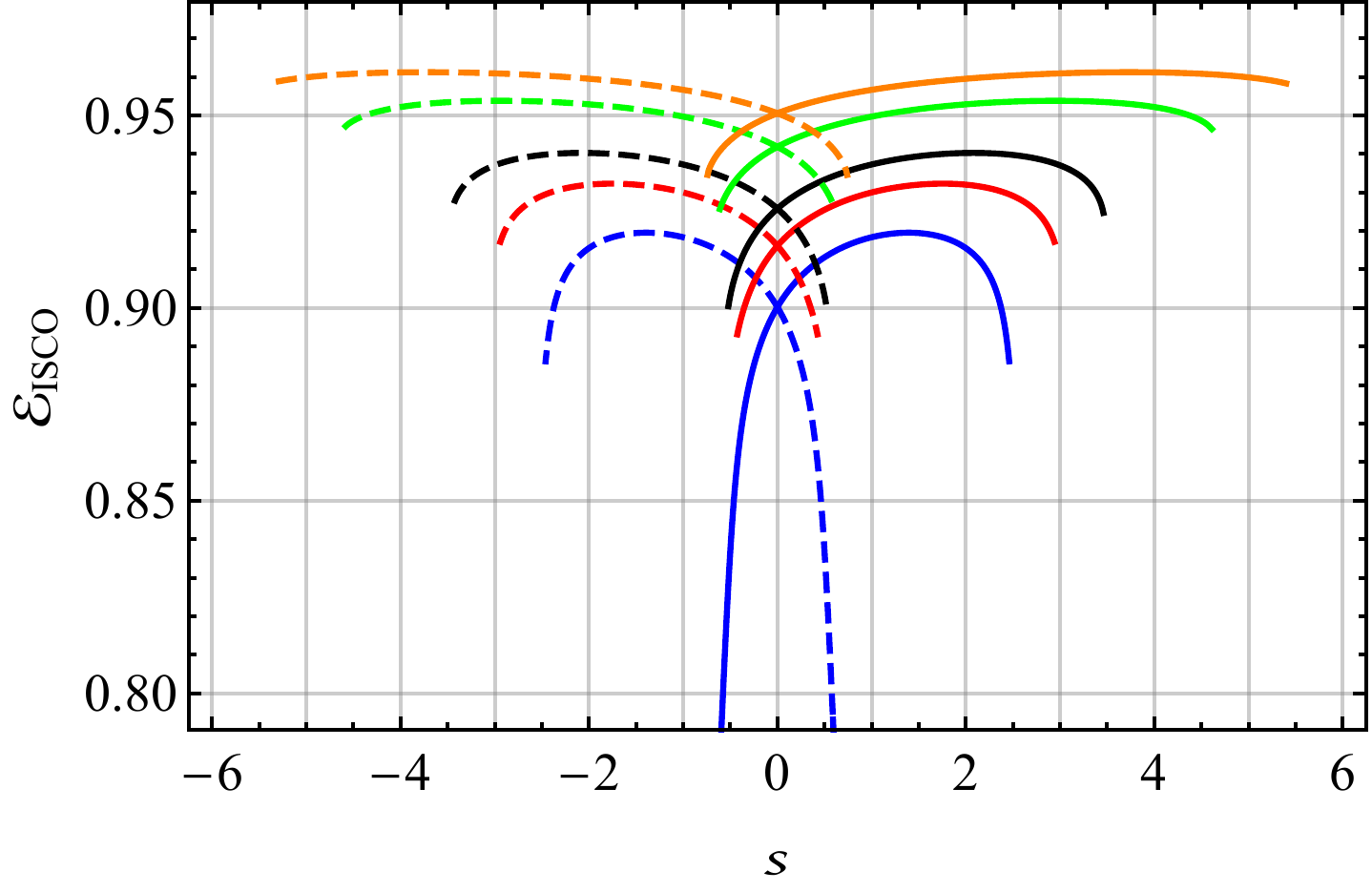}&\includegraphics[scale=0.55]{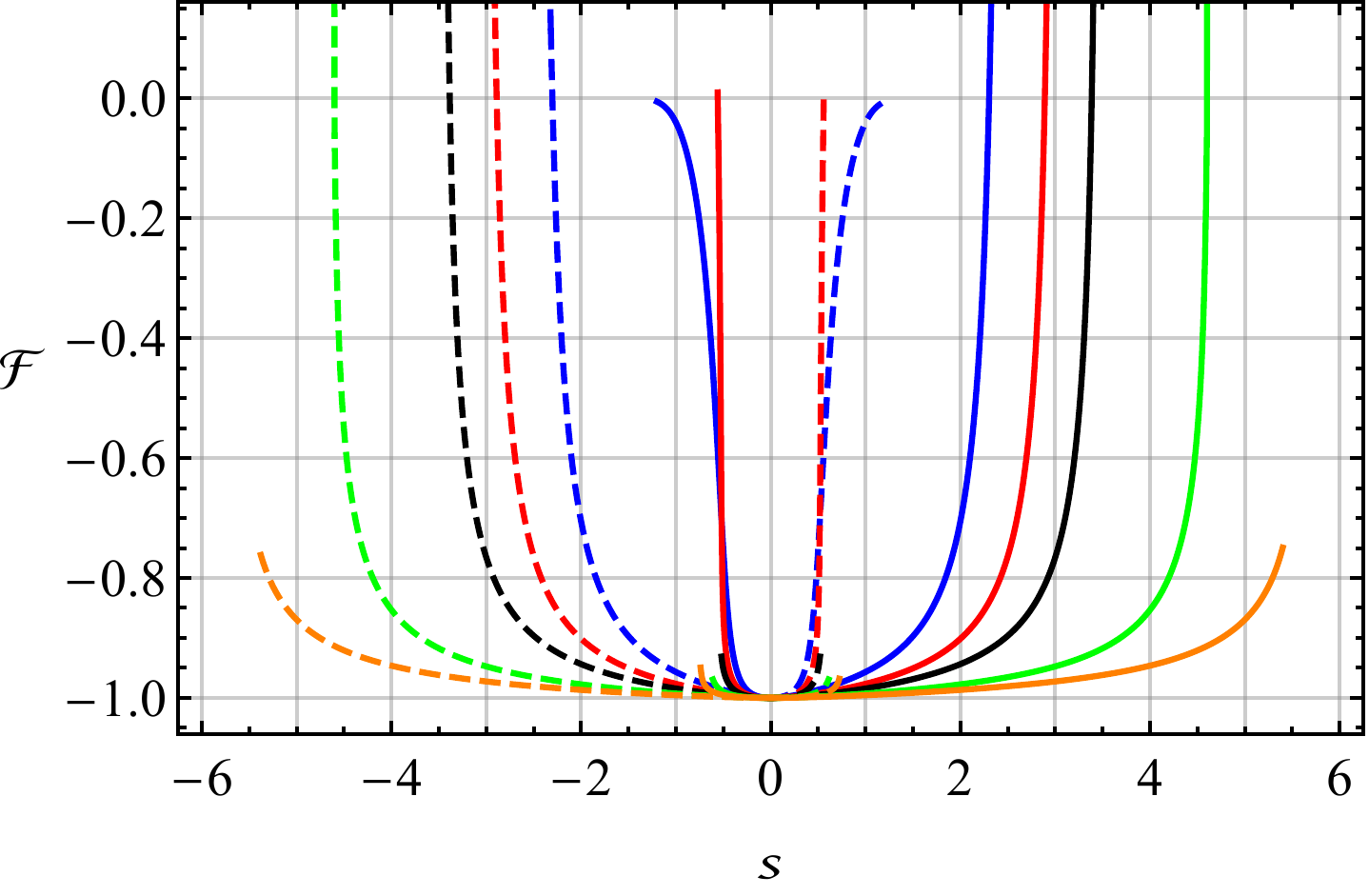}
    \end{array}$
    \end{center}
    \caption{$l_{ISCO}$, $\mathcal{L}_{ISCO}$, $\mathcal{E}_{ISCO}$ and the function $\mathcal{F}$ vs. $s$ for different values of $a=-0.1$ (blue), $a=-0.2$ (red), $a=-0.3$ (black), $a= -0.6$ (green) and $a=-0.9$ (orange) in the upper (continuous line) and lower (dashed line) universes. We assume $b_0=M=1$. \label{fig7a}}
    \end{figure*}
    %%%%%%%%%%%%%%%%%

	In Fig.\ref{fig4}, we plot $\mathcal{V}_{\text{eff}}$ vs. $l$ for different values of $a$, $s$, and $\mathcal{L}$. In the left panel, we show the behavior of the effective potential for different values of $a$. The figure shows how the effective potential decreases as the wormhole's spin $a$ increases. In this case, the minimum value for the effective potential decreases from $\mathcal{V}_\text{min}\approx0.5$ to $\mathcal{V}_\text{min}\approx-1$ as the wormhole's spin $a$ changes from $0.1$ to $0.4$. In the central panel, we show the behavior of $\mathcal{V}_\text{eff}$ for different values of $s$. From the figure, it is possible to see that $\mathcal{V}_\text{eff}$ does not change significantly as $s$ increases from $0.1$ to $0.4$. The figure also shows that $\mathcal{V}_\text{eff}$ in the lower universe is larger than those in the upper universe for all values of $s$. Furthermore, the value of the $\mathcal {V}_\text{eff}$ at the throat is the same for all values of $s$; in contrast to the first and third panels, where its value decreases as $a$ increases and decreases as $\mathcal{L}$ increases, respectively.

	In the right panel of Fig.~\ref{fig4}, we plot the effective potential as a function of $l$ for different values of the angular momentum $\mathcal{L}$ while keeping constant $a$ and $s$. From the figure, we can see how $\mathcal{V}_\text{eff}$ decreases as the particle's angular momentum $\mathcal{L}$ increases. The figure also shows that the effective potential is always larger in the lower universe than in the upper one.

    \subsection{Innermost stable circular orbits}

	Now, we focus our attention on circular orbits of spinning test particles in the geometry of a rotating wormhole given by Eq.~(\ref{s4ae3}). It is well-known that circular orbits occur when the radius of an orbit of particles is constant. Therefore, the radial velocity of a spinning test particle vanishes $dl/d\lambda=0$. In this sense, according to Eq.~(\ref{s3ae10}),   $\mathcal{E}=\mathcal{V}_{\text{eff}}$, where we have defined the energy of the test particle per unit mass as $\mathcal{E}=E/m$. On the other hand, the radial acceleration of the particle also vanishes, i.e. $d^2l/d\lambda^2=0$, which implies that $d\mathcal{V}_\text{eff}/dl=0$. Nevertheless, these conditions do not guarantee the stability of circular orbits. To ensure stability, the second radial derivative of the effective potential must be positive, namely,
    \begin{equation}
    \label{4.1}
    \frac{d^2\mathcal{V}_\text{eff}}{dl^2}\geq 0.
    \end{equation} 
    %%%%%%%%%%%%%%%%%%%%%%%%%%%%%%%%%%%%%%%%%%%%%%%%%%%%%%%%%%
    \begin{table*}[t]
    \caption{\label{table1}
    Values for $l_{\text{ISCO}}$, $\mathcal{L}_{\text{ISCO}}$ and $\mathcal{E}_\text{ISCO}$ for different values of $a$ and $s$. Here, $U$ and $L$ subscripts denote upper and lower universes, respectively.}
    \begin{ruledtabular}
    \begin{tabular}{cccccccc}
    $a$&$s$&$l_{\text{ISCO}}$ &$\mathcal{L}_{\text{ISCO}}$&$\mathcal{E}_{\text{ISCO}}$&$l_{\text{ISCO}}$&$\mathcal{L}_{\text{ISCO}}$&$\mathcal{E}_{\text{ISCO}}$\\
    \hline
    0.1 & 0.0 & $\mp$ 2.994 & -2.403 & 0.900   & --- & --- & ---\\      
    				&   $\pm 0.1$    &    $\mp$3.155    &   -2.356     &   0.904    &   $\pm$2.803    &   -2.440     &   0.895   \\
    				&  $ \pm$ 0.4    & $ \mp$ 3.519    &   -2.180     &   0.911    &  $\pm$ 1.901     &  -2.436     & 0.865    \\
    				&  $\pm$ 0.8    &  $\mp$ 3.816  &   -1.893     &   0.917    & $ \pm$ 0.895  &   -1.649    & 0.682 \\
    				&  $\pm$ 1.31  & $ \mp$ 3.962   &  -1.472   & 0.919   &  $\pm$ 0.601     &  -0.874 &  0.502   \\
    				& $\pm$ 1.6  & $\mp$ 3.925 & -1.210  & 0.919 &   ---  &  ---  &  ---   \\
    				&  $\pm$ 2.0  & $\mp$ 3.663  &  -0.814  &  0.915 & --- &  ---  &  --- \\
    				&   &  &   &   &  &    &   \\
    				%&  $\pm$ 2.3   &  $\mp$ 3.107  & -0.467  &  0.905 & ---  &  ---  &  --- \\
    0.2& 0.0 & $\mp$ 3.714 &  -2.635  & 0.916  & ---  & --- & ---  \\  
    				&  $\pm$ 0.1  &  $\mp$ 3.870  &  -2.581  &  0.919  &  $\pm$ 3.534 &  -2.682  & 0.913 \\
    				& $\pm$ 0.55  &  $\mp$ 4.376  &  -2.289 &  0.926 & $\pm$ 1.209 & -2.496  & 0.840  \\
    				& $\pm$ 1.0 & $\mp$ 4.675 & -1.946  & 0.930  &  --- & ---  &  --- \\
    				& $\pm$ 1.5  & $\mp$ 4.829 &  -1.526 &  0.932 & --- &  --- &  --- \\ 
    				&  $\pm$ 2.0 &  $\mp$ 4.796 & -1.073 &  0.932 &  ---   &  --- &  --- \\  
    				&  $\pm$ 2.5 &  $\mp$ 4.482 & -0.581 &  0.929 &  ---   &  --- &  --- \\ 
    				&   &  &   &   &  &    &   \\
    				%&  $\pm$ 2.89 &  $\mp$ 3.685 & -0.139 &  0.920 &  ---   &  --- &  --- \\   
    0.3& 0.0 &  $\mp$ 4.285  & -2.813   & 0.926  & --- & ---   & --- \\ 
    				&  $\pm$ 0.1 &  $\mp$ 4.439 & -2.755  &  0.928  & $\pm$ 4.109 &  -2.865  & 0.923 \\
    				&  $\pm$ 0.52  & $\mp$  4.928 & -2.476   &  0.933 &  $\pm$ 2.639 &  -2.948  &  0.900 \\
    				&  $\pm$ 1.2 &  $\mp$ 5.388 &  -1.948  & 0.938 &  ---  &  --- & --- \\
    				&  $\pm$ 1.8 &  $\mp$ 5.555 & -1.435  & 0.940 &  ---  &  --- & --- \\
    				&  $\pm$ 2.4 &  $\mp$ 5.497 &  -0.884  & 0.940 &  ---  &  --- & --- \\
    				&  $\pm$ 3.0 &  $\mp$ 5.076 &  -0.290  & 0.937 &  ---  &  --- & --- \\
    				&   &  &   &   &  &    &   \\
    				%&  $\pm$ 3.38 &  $\mp$ 4.176 &  0.138  & 0.930 &  ---  &  --- & --- \\
    0.6& 0.0 &  $\mp$ 5.619 & -3.204  & 0.942 & --- &  --- &  ---  \\
    				&  $\pm$ 0.1 &  $\mp$ 5.758 & -3.147  &  0.943  & $\pm$ 5.448 &  -3.264  & 0.940 \\
    				&  $\pm$ 0.61 &  $\mp$ 6.377 & -2.782  &  0.947  & $\pm$ 3.826 &  -3.456  & 0.925 \\
    				&  $\pm$ 1.6 &  $\mp$ 7.055 &  -1.987 &  0.952  & ---  & --- &  --- \\
    				&  $\pm$ 2.4 &  $\mp$ 7.290 &  -1.289 &  0.953  & ---  & --- &  --- \\
    				&  $\pm$ 3.2 &  $\mp$ 7.259 &  -0.554 &  0.954  & ---  & --- &  --- \\
    				&  $\pm$ 4.0 &  $\mp$ 6.818 &  0.224 &  0.952  & ---  & --- &  --- \\
    				&   &  &   &   &  &    &   \\
    				%&  $\pm$ 4.61 &  $\mp$ 4.670 &  0.880 &  0.946  & ---  & --- &  --- \\
    0.9& 0.0 & $\mp$ 6.684 &  -3.494 &  0.951 &  --- &  --- & --- \\ 
    				& $\pm$ 0.1  &  $\mp$ 7.598  & -2.961 &  0.955 & $\pm$ 6.515 & -3.558  &  0.950 \\
    				& $\pm$ 0.74  &  $\mp$ 6.839  & -3.428 &  0.951 & $\pm$ 4.051 & -3.815  &  0.935 \\
    				&  $\pm$ 1.8 &  $\mp$ 8.351 &  -2.098 &  0.959  & ---  & --- &  --- \\
    				&  $\pm$ 2.7 &  $\mp$ 8.677 &  -1.312 &  0.961  & ---  & --- &  --- \\
    				&  $\pm$ 3.6 &  $\mp$ 8.749 &  -0.492 &  0.961  & ---  & --- &  --- \\
    				&  $\pm$ 4.5 &  $\mp$ 8.498 &  0.362 &  0.961  & ---  & --- &  --- \\
    				&   &  &   &   &  &    &   \\
    				%&  $\pm$ 5.4 &  $\mp$ 7.466 &  1.265 &  0.958  & ---  & --- &  --- \\
    \end{tabular}
    \end{ruledtabular}
    \end{table*}
    %%%%%%%%%%%%%%%%%%%%%%%%%%%%%%%%%%%%%%%%%%%%%%%%%%

    Hence, the marginally stable circular orbit, corresponding to the smallest allowed value for stable circular orbits, also known as the innermost stable circular orbit or ISCO, can be obtained when $d^2\mathcal{V}_\text{eff}/dl^2=0$.

    Using the conditions $\mathcal{E}=\mathcal{V}_{\text{eff}}$ and $d\mathcal{V}_\text{eff}/dl=0$ we obtain the values of $\mathcal{E}$ and $\mathcal{L}$ in terms of the circular orbit radius $l$ and then from $d^2\mathcal{V}_\text{eff}/dl^2=0$ the value of the radius of the ISCO for a spinning test particle. Since the process involves the solution of a non-linear system of equations for $l$ and $\mathcal{L}$, we must solve it numerically, see table~\ref{table1}.

	In Fig.~(\ref{fig6}), we show the first and second derivatives of $\mathcal{V}_\text{eff}$ as a function of $l$. The intersection between the curves correspond to the innermost stable circular orbit  $l_{ISCO}$. According to the figure, when $a=0.1$ and $s=1.3$, there are two possible values for $l_{ISCO}$. One in the lower universe $l^L_{\text{ISCO}}=-3.961$, and one in the upper universe $l^U_\text{ISCO}=0.607$\footnote{We use the upscript $L$ and $U$ for lower and upper universes, respectively.}.  

	In Figs.~\ref{fig7} and \ref{fig7a},  we plot the behavior of the radius (top panel left), angular momentum (top panel right), energy (lower panel left), and  $\mathcal{F}$ (lower panel right)  at the ISCO for different values of the wormhole spin $a$, co-rotating ($a>0$) and counter-rotating ($a<0$), respectively\footnote{The behavior between co-rotating and counter-rotating is similar. For this reason, we focus our discussion to the former case.}. In Fig.~\ref{fig7}, we can see how $l_{ISCO}$ increases as $a$ increases. For example, if we consider the upper universe, the values of the ISCO for $a=0.9$ (orange) are larger than those with $a=0.1$. In the lower universe, on the other hand, $l_{ISCO}$ becomes smaller as the spin parameter $a$ increases. See the dashed lines. 

	For constant values of $a$, the figure shows how $l^U_{ISCO}$ increases as $s$ increases, reaching a maximum value. Then, the ISCO radius decreases as $s$ increases; see the continuous lines in the top-left panel of Fig.~\ref{fig7}. In the lower universe, the situation is the opposite: $l_{ISCO}$ decreases until it reaches a minimum value, and then it increases as $s$ increases. For example, when $a=0.2$, the maximum/minimum value for $l_{ISCO}$  is around $5$ and $-5$, respectively. While for $a=0.9$, the maximum/minimum value for $l_{ISCO}$  is around $6$ and $-6$, respectively. The figure also shows that there is an interval in which there is only one value of $l_{ISCO}$, and one interval in which there are two values. For example, in the upper universe, when $a=0.1$, the figure shows only one value of $l_{ISCO}$ if $-2.5\geq s< -1.25$. When $|s| \leq 1.25$, the figure shows two values: one in the upper universe and the other in the lower universe. Then, when $s>1.25$, the figure shows only one value of $l_{ISCO}$ located in the lower universe. One can see a similar behavior for other values of $a$ with different ranges for $s$. 

\begin{figure}[t]
    \begin{center}
    \includegraphics[scale=0.55]{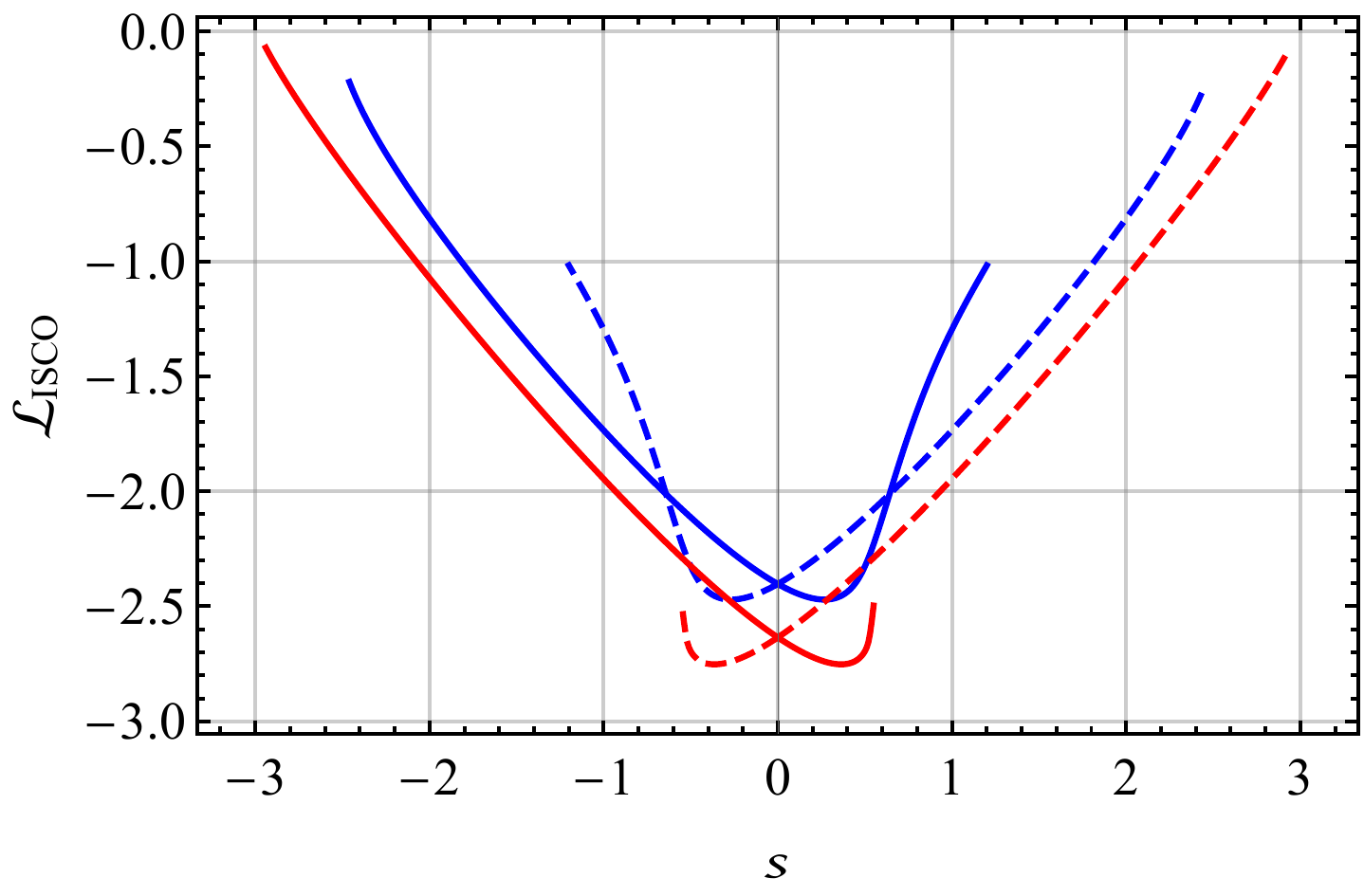}				
    \end{center}
    \caption{$\mathcal{L}_{ISCO}$ vs. $s$ for $a = 0.1$ (blue) and $0.2$ (red). We assume $b_0=M=1$. \label{fig8}}
    \end{figure}

	In the top-right panel of Fig.~\ref{fig7}, we show the behavior of $\mathcal{L}_{ISCO}$ as a function of $s$ for different values of $a$. According to the figure, the value of $\mathcal{L}_{ISCO}$ in the upper universe decreases as $s$ increases. Nevertheless, for values of $a<0.3$, $\mathcal{L}_{ISCO}$ reaches a minimum value and starts to increase again, see the blue and red (continuous) lines. On the other hand, when we consider the lower universe (dashed lines), the behavior is the contrary. For values of $a\geq 0.3$, $\mathcal{L}_{ISCO}$ decreases as $s$ increases. While for values of $a<0.3$, $\mathcal{L}_{ISCO}$ decreases, reaches a minimum, and starts to increase again, see the dashed blue and red lines in Fig.~\ref{fig8}.

    %%%%%%%%%%%%%%%%%%%%
    %%%%%%%%%%%%%%%%%%%%

	In the bottom-left panel of Fig.~\ref{fig7}, we plot the behavior of the energy at the ISCO for different values of the wormhole's spin $a$. From the figure, for constant values of the spin $s$, we see that the energy $\mathcal{E}_{ISCO}$ increases as $a$ increases. Moreover, the figure also shows that $\mathcal{E}_{ISCO}$ is always lower than $\mathcal{E}=1$. In the upper universe, while $s$ increases, $\mathcal{E}_{ISCO}$ increases. Then, it reaches a maximum value and decreases until some value of $s$, the lowest limit for the particle's spin. On the other hand, when we consider the lower universe, the behavior is the opposite (as before). $\mathcal{E}_{ISCO}$ increases as $s$ increases. Then, it reaches a maximum value and decreases until some value of $s$.

    \subsection{Superluminal bound}

	As mentioned before, we can obtain the superluminal bound using Eq.~(\ref{s3be1}). In this sense, after replacing $dr/dt$ and $d\varphi/dt$ (see Eq.~\ref{s3be10}), we obtain
    \begin{equation}
    \begin{aligned}
    \label{s4ce1}
    \mathcal{F}&=g_{tt}(\hat{\mathcal{B}}p_t+\hat{\mathcal{D}}p_\varphi)^2+g_{rr}(\hat{\mathcal{C}}p_r)^2+2g_{t\varphi}(\hat{\mathcal{D}}p_t+\hat{\mathcal{A}}p_\varphi)\\
    &\times(\hat{\mathcal{B}}p_t+\hat{\mathcal{D}}p_\varphi)+g_{\varphi\varphi}(\hat{\mathcal{D}}p_t+\hat{\mathcal{A}}p_\varphi)^2\leq 0.
    \end{aligned}
    \end{equation}
    Therefore, the limit value of $s$ for which the spinning test particle's motion is time-like can be obtained with the condition $\mathcal{F}=0$. Then, using Eqs.~(\ref{s3ae8}) for $p_r$ and replacing into Eq.~(\ref{s4ce1}), we get  
    \begin{equation}
    \label{s4ce1a}
    \mathcal{F}= \mathcal{X}_1\left(\frac{p_t}{m}\right)^2+ 2\mathcal{X}_2\left(\frac{p_t}{m}\right)\left(\frac{p_{\varphi}}{m}\right)+\mathcal{X}_3\left(\frac{p_{\varphi}}{m}\right)^2- \mathcal{X}_4 \leq 0,
    \end{equation}
    with\footnote{Here, we keep the notation of Eq.~(\ref{s3ae1}), where $r$ is the radial coordinate. Nevertheless, in the rotating wormhole, the radial coordinate changes to $l$, so $r\rightarrow l$ in all the equations.}
    \begin{equation}
    \label{s4ce1b}
    \begin{aligned}
    \mathcal{X}_1&= g_{tt}\mathcal{\hat{B}}^2+2g_{t\varphi}\mathcal{\hat{B}}\mathcal{\hat{D}}+g_{t\varphi}\mathcal{\hat{D}}^2-g^{tt}g_{rr}^2\mathcal{\hat{C}}^2\\\\
    \mathcal{X}_2&= g_{tt}\mathcal{\hat{B}}\mathcal{\hat{D}}+g_{t\varphi}(\mathcal{\hat{A}}\mathcal{\hat{B}}+\mathcal{\hat{D}}^2)
    +g_{\varphi\varphi}\mathcal{\hat{A}}\mathcal{\hat{D}}-g^{t\varphi}g_{rr}^2\mathcal{\hat{C}}^2 \\\\
    \mathcal{X}_3&= g_{tt}\mathcal{\hat{D}}^2+2g_{t\varphi}\mathcal{\hat{A}}\mathcal{\hat{D}}+g_{\varphi\varphi}\mathcal{\hat{A}}^2-g^{\varphi\varphi}g_{rr}^2\mathcal{\hat{C}}^2\\\\
   \mathcal{X}_4 &= g_{rr}^2\mathcal{\hat{C}}^2.
    \end{aligned}
    \end{equation}
    Note that Eq.~(\ref{s4ce1a}) reduces to Eqs.~(45) of Ref.~\cite{Benavides-Gallego:2021lqn} when we consider the space-time of a non-rotating wormhole.
    
    %%%%%%%%%%%%%%%%%%%%%%%%
    \begin{table}[t]
    \caption{\label{table2}
    Superluminal bound for upper and lower universes in the co-rotating case ($a>0$). In the table $s_+$ and $s_-$ are the limits for positive and negative values of $s$, respectively.}
    \begin{ruledtabular}
    \begin{tabular}{cccccc}
    $a$&Universe&$s_+$&$s_-$ &$\mathcal{F}_+$&$\mathcal{F}_-$\\
    \hline\\
    0.1  & \text{Upper} & 1.315 & -2.30305 & $-4.8\times10^{-6}$& $-4.1\times10^{-6}$ \\  
    & \text{Lower} & 2.30305 & -1.315 & $-4.1\times10^{-6}$& $-4.8\times10^{-6}$\\  
    0.2 & \text{Upper} & 0.558933 & -2.89066 & $-4.2\times10^{-7} $ & $-6.3\times10^{-6} $\\  
    & \text{Lower} & 2.89066 & -0.558933 & $-6.3\times10^{-6}$ &  $-4.2\times10^{-7} $  \\ 
    0.3 & \text{Upper} & 0.52 & -3.38559 & -0.931192& $-4.5\times10^{-6}$ \\  
    & \text{Lower} & 3.38559 & -0.52 & $-4.5\times10^{-6}$& -0.931192\\   
    0.6 & \text{Upper} & 0.61 & -4.60353 & -0.9693& $-3.0\times10^{-6}$\\  
    & \text{Lower} & 4.60353  & -0.61 & $-3.0\times10^{-6}$& -0.9693\\  
    0.9 & \text{Upper} & 0.74 & -5.4 & -0.9495 & -0.7493\\  
    & \text{Lower} & 5.4 & -0.74 & -0.7493& -0.9495\\          
    \end{tabular}
    \end{ruledtabular}
    \end{table}

	In the bottom-left panel of Fig.~(\ref{fig7}), we plot the behavior of $\mathcal{F}$ as a function of $s$ for different values of $a$. From the figure, it is possible to identify an interval for the particle's spin $s$ in both the upper and lower universes. There are always two limits, one for negative values of the spin $s_-$ and the other for positive values, $s_+$. For example, in the upper universe, when the value of $a=0.1$, the function $\mathcal{F}$ reaches the superluminal bond at $s_-\approx -2.3$ and $s_+\approx 1.3$, see the continuous blue line. One can see similar behavior when $a=0.2$. In that case, $\mathcal{F}$ reaches the superluminal bound when $s_- \approx -2.8$ and $s_+\approx 0.5$. Finally, when $a=0.3$, the spinning test particle reaches the superluminal bound when $s_-\approx -3.3$. Nevertheless, when $s>0.52$, the system of non-linear equations 
    \begin{equation}
    \label{s4ce2}
    \begin{array}{ccc}
    d\mathcal{V}_\text{eff}/dl=0&\text{and}&d^2\mathcal{V}_\text{eff}/dl^2=0.\end{array}
    \end{equation} 
    does not have a solution. As a consequence, $\mathcal{F}$ does not reach the superluminal bound $\mathcal{F}=0$. However, although $\mathcal{F}<0$, we can consider $s\approx0.52$ as the positive limit value for a spinning test particle in the upper universe, $s_+$. 

	On the other hand, for large values of $a$ (e. i. 0.9), the figure shows that $\mathcal{F}$ does not reach the superluminal limit neither for negative nor for positives values of $s$. Once again, although $\mathcal{F}$ is always negative in these cases, we can use these values to set an interval of motion for a spinning test particle. Hence, when $a=0.9$ the limit values are $s_-\approx -5.4$ and $s_+\approx 0.74$. 

	When we consider the lower universe (dashed lines in Fig.~\ref{fig7}), the figure shows the same behavior as in the upper universe. Once again, $\mathcal{F}$ gives two limit values for $s<0$ and $s>0$. Nevertheless, due to symmetries in $\mathcal{V}_{\text{eff}}$, these values are the same and only change in the sign. For example, in the upper universe, we found that  $s_+\approx1.3$ and $s_-\approx-2.3$ when $a=0.1$. Therefore, due to the symmetries, in the lower universe $s_-\approx-1.3$ while $s_+\approx2.3$. In table~\ref{table2}, we show some of the values. 	
			
    \section{Conclusions \label{SecV}}
        
    In this work, we investigated the motion of spinning test particles around a rotating wormhole, an extension of the previous work presented by Benavides-Gallego et al. in Ref.~\cite{Benavides-Gallego:2021lqn}, where the authors take into account a spinning test particle moving around a non-rotating traversable wormhole, the well-known Morris-Thorne wormhole~\cite{Morris:1988cz}. There, the authors showed how the effective potential depends on the wormhole's spin $a$, the particle's spin $s$, the proper distance $l$, the angular momentum $\mathcal{L}$, and the wormhole's throat $b_0$. According to the authors, the effective potential $\mathcal{V}_{\text{eff}}$ shows some symmetries in its behavior, represented by two configurations: \textit{plus} and \textit{minus}. In the plus configuration, $\mathcal{V}^P_{\text{eff}}(l,s,\mathcal{L})=\mathcal{V}^P_{\text{eff}}(l,-s,-\mathcal{L})$. In the minus configuration, on the other hand, we have that $\mathcal{V}^M_{\text{eff}}(l,-s,\mathcal{L})=\mathcal{V}^M_{\text{eff}}(l,s,-\mathcal{L})$. Therefore, a spinning test particle moving with clockwise spin and angular momentum is equivalent to a particle moving with counterclockwise spin and angular momentum (\textit{plus} configuration); while a spinning test particle moving with clockwise spin and counterclockwise angular momentum is equivalent to a particle moving with counterclockwise spin and clockwise angular momentum (\textit{minus} configuration).
        
    In the rotating case, we also found that $\mathcal{V}_{\text{eff}}$ has ``\textit{plus}'' and ``\textit{minus}'' configurations defined as $\mathcal{V}^P_{\text{eff}}(l,s,a,\mathcal{L})$ and $\mathcal{V}^P_{\text{eff}}(l,-s,a,\mathcal{L})$, respectively. Moreover, since $\mathcal{V}_{\text{eff}}$ depends on the wormhole's spin $a$, there are more relations of symmetry, see footnote $10$. In each configuration, the ISCO has the same value. However, if we change the configuration, the ISCO also changes. Hence, if $l_{ISCO}$ is in the upper universe for the ``\textit{plus}'' configuration, it will change to the lower universe once we chance to the ``\textit{minus}'' configuration, see Fig.~\ref{fig1}. Furthermore, from the dynamical point of view, it is important to point out that these symmetries allow some equivalences in the motion of spinning test particles regarding the sign of $s$, $\mathcal{L}$ and $a$.
   		
    On the other hand, the shape of $\mathcal{V}_{\text{eff}}$ is symmetric only in two particular cases: $s=a=0$ or $s=0$ and $a\neq0$. In these situations, considering the upper and lower universes, the ISCO is at the same distance from the throat. Moreover, $\mathcal{V}_{\text{eff}}$ reaches a minimum value at $l = 0$. On the other hand, when $a$ and $s$ are different from zero, $\mathcal{V}_{\text{eff}}$ is non-symmetric and looks like the effective potential of a spinning test particle in a non-rotating wormhole. Nevertheless, in contrast with the non-rotating case, where the effective potential reaches a minimum value at the throat ($l= 0$), $\mathcal{V}_{\text{eff}}$ has a minimum value shifted to the right or left depending on the configuration, see Fig.~\ref{fig1}. It is important to point out that $\mathcal{V}_\text{min}$ decreases in three different situations: when the wormhole's spin, the particle's spin or angular momentum increases while keeping constant $s$ and $\mathcal{L}$, $a$ and $\mathcal{L}$, and $a$ and $s$, respectively. See Fig.~\ref{fig4}. 
    
    In the non-rotating case, Ref.~\cite{Benavides-Gallego:2021lqn} shows that there is only one value for $l_{ISCO}$ (in the upper or lower universes) when $|s|\geq 1$ and two values when $-1<s<1$ (one in the upper universe and the other in the lower universe), see Fig. 8 in Ref.~\cite{Benavides-Gallego:2021lqn}. We found a similar behavior for the rotating case. However, the interval for $s$ in which there are two values of $l_{ISCO}$ changes as the wormhole's spin $a$ increases. 
   
    If we consider the lower universe, the analysis performed in Ref.~\cite{Benavides-Gallego:2021lqn} shows that $l_{ISCO}$ decreases as $s$ increases. Then, it reaches a minimum value at $s\approx -0.8$ and increases again up to the throat (when $s=1$), see Fig.~8 in Ref.~\cite{Benavides-Gallego:2021lqn}. The opposite behavior occurs when we consider the upper universe. In the rotating case, on the other hand, our analysis shows that $l_{ISCO}$ in the lower universe decreases as $s$ decreases. Then, it reaches a minimum value and increases again. Nevertheless, in contrast to the non-rotating case, $l_{ISCO}$ does not arrive at the wormhole's throat. See the upper-left panel of Fig.~\ref{fig7}, \textcolor{black}{This behavior may be a consequence of the shifting effect on $\mathcal{V}_{\text{eff}}$ at the minimum.} 
   
    In Ref.~\cite{Zhang:2017nhl}, Y.~P.~Zhang et al. investigated the ISCO orbit for a classical spinning test particle in the background of a Kerr-Newman black hole. There, the authors show that the motion of the spinning test particle is related to its spin, and it will be superluminal if its spin is large. According to their analysis, the authors found that $l_{ISCO}$ decreases as the spin of the particle increases. We also found the same behavior in the rotating wormhole when the upper universe is considered (the behavior is the opposite in the lower universe). Nevertheless, in contrast with the results of Ref.~\cite{Zhang:2017nhl}, where a spinning test particle can orbit in a smaller circular orbit than a non-spinning test particle, we found that $|l_{ISCO}|$ increases as the wormhole's spin $a$ increases (while keeping constant the particles' spin $s$). In this sense, a spinning test particle can move in larger circular orbits than the non-rotating case.
   
    In this work, we found that the energy of the innermost stable circular orbit $\mathcal{E}_{ISCO}$ is always below unity, i.e. $\mathcal{E}_{ISCO}< 1$. This result agrees with the non-rotating case, where $\mathcal{E}_{ISCO}<0.9$~\cite{Benavides-Gallego:2021lqn}. Furthermore, it is important to remark that the energy increases as the wormhole's spin $a$ increases. Hence, $\mathcal{E}_{ISCO}$ in the non-rotating case is smaller than the energy in the rotating case, where $0.8<\mathcal{E}_{ISCO}<0.96$. In Ref.~\cite{Benavides-Gallego:2021lqn}, the authors found that $\mathcal{E}_{ISCO}$ has the same value for both the lower and upper universes when $s=0$. In our analysis, we found the same behavior, which is a consequence of the mirror-like symmetric of $\mathcal{V}_\text{eff}$\footnote{Recall that $\mathcal{V}_{\text{eff}}$ is symmetric when $s=0$ in both the rotating and non-rotating wormhole, see the first and third panels in Fig~\ref{fig2}.}. 
    
    In this work, we also found that the angular momentum at the ISCO behaves differently from the non-rotating case. In Ref.~\cite{Benavides-Gallego:2021lqn}, the authors showed that the angular momentum reaches the maximum value $\mathcal{L}_{ISCO}\approx2.002$ when $|s|\approx 0.25$~\cite{Benavides-Gallego:2021lqn} and it is always positive, $\mathcal{L}_{ISCO}>0$. Moreover, in the lower universe, the angular momentum increases as $s$ increases, reaching its maximum value at $s\approx -0.25$. Then, it decreases for $-0.25<s<2$. See the left-bottom panel of Fig.~8 in Ref.~\cite{Benavides-Gallego:2021lqn}. The opposite behavior occurs when we consider the upper universe. In the rotating case, on the other hand, we found that $-4<\mathcal{L}_{ISCO}<1.4$. Furthermore, for small values of $s$ ($0.1$, $0.2$ or $0.3$), we show $\mathcal{L}_{ISCO}$ in the upper universe decreases as $s$ increases. Then, it reaches a minimum value and starts to increase again to a certain value of $s$. The opposite behavior occurs in the lower universe. Furthermore, for larger values of $s$ ($0.6$ or $0.9$), the $\mathcal{L}_{ISCO}$ in the lower universe decreases as $s$ decreases, reaching a minimum value.
   
    Finally, we consider the superluminal bound to investigate the constraints for the particle's spin $s$. In the non-rotating case, Benavides-Gallego et al. showed that the motion of a spinning test particle has physical meaning (the trajectory is time-like) if $-1.5<s<1.5$. Nevertheless, for values of $|s|>1.5$, the particle's trajectory is superluminal (space-like) and it does not have a physical meaning. In the rotating case, the shape of $\mathcal{F}$ is similar to the non-rotating case; but we found some differences. For example, when one considers small values of $a$ ($0.1$, or $0.2$), $\mathcal{F}=0$ for two values of the particle's spin: $s_-$ and $s_+$; see the right-bottom panel of Fig.~\ref{fig7}. Moreover, for $|s|<1$ and $a=0.3$, the function $\mathcal{F}$ stops because the non-linear system in Eq.~(\ref{s4ce2}) does not have a solution, setting a constraint value for $s$ which is positive/negative for the upper/lower universe, see table~\ref{table2}. For larger values of $a$ ($0.9$), the function $\mathcal{F}$ never reaches the superluminal bound ($\mathcal{F}=0$) because the non-linear system Eq.~(\ref{s4ce2}) does not have solution. In this sense, the spin is constrained to those values for which the non-linear system has a solution, see table~\ref{table2}. 
   
    It is important to remark that our analysis uses the MPD equations. Therefore,  we considered the approximation in which the mass and size of the spinning test particle are negligible in relation to the mass of the central object and must not affect the geometry background. Nevertheless, from the astrophysical point of view, the motion of spinning test particles may still determine some features that enable us to distinguish black holes from wormholes. As we have shown in this paper, the spin does affect the motion of test particles around a rotating wormhole. On the other hand, observationally speaking, the spinning test particles may form the accretion disk of black holes. These particles could be larger objects, such as asteroids, planets/exoplanets orbiting stellar-mass objects, or rapidly rotating black holes and neutron stars orbiting supermassive candidates. Hence, the spin could be a crucial parameter to consider when describing the motion of such objects and both the electromagnetic and gravitational wave observations that would allow us to conclude if these objects are black holes or wormholes.

%%%%%%%%%%%%%%%%%%%%%%%%%%%%%%%%%%%%%%%%%%%%%%%%%%%%%%%%%%%%%%%%%%%%%%%%%%%%%%%%%% Acknowledgments %%%%%%%%%%%%%%%%%%%%%%%%%
%%%%%%%%%%%%%%%%%%%%%%%%%%%%%%%%%%%%%%%%%%%%%%%%%%%%%%%%%%%%%%   
   %\newpage
   \begin{acknowledgments}
    The work of F.~A and C.~A.~B.~G is supported, respectively, by Grant F-FA-2021-432 of the Uzbekistan Ministry for Innovative Development and the PIFI program of the Chinese Academy of Sciences. W. H. is supported by CAS Project for Young Scientists in Basic Research YSBR-006, and acknowledges the support of The National Key R\&D Program of China (Grant No. 2021YFC2203002), NSFC (National Natural Science Foundation of China) No. 12173071 and No. 11773059. J.R, F.A. and A.A. acknowledges the support of the grants F-FA-2021-432, F-FA-2021-510, and MRB-2021-527 of the Uzbekistan Ministry for Innovative Development. A.A. also acknowledges the support of the PIFI program and J.R. thanks to the ERASMUS+ project 608715-EPP-1-2019-1-UZ-EPPKA2-JP (SPACECOM).  
    %%%%
   \end{acknowledgments}

%%%%%%%%%%%%%%%%%%%%%%%%%%%%%%%%%%%%%%%%%%%%%%%%%%%%%%%%%%%%%%%%%%%%%%%%%%%%% Appendix A %%%%%%%%%%%%%%%%%%%%%%%%%%%%%%%%%%%
%%%%%%%%%%%%%%%%%%%%%%%%%%%%%%%%%%%%%%%%%%%%%%%%%%%%%%%%%%%%%%
	\newpage
	\appendix
	\section{The dragging of the inertial frame\label{appenA}}
	In this appendix, we follow Chandrasekhar  to show the existence of a dragging effect in the line element given by Eq.~(\ref{s2be1}). Hence, the contravariant components of the metric tensor $g^{\mu\nu}$ can be expressed in the matrix form as\footnote{Recall that we set $t\rightarrow 0$, $\theta \rightarrow 2$ and $r\rightarrow 3$.}
	\begin{equation}
	    \label{A1}
	    (g^{\mu\nu})=\left(
        \begin{array}{cccc}
        -\frac{1}{N^2} & -\frac{\omega }{N^2} & 0 & 0 \\
        -\frac{\omega }{N^2} & -\frac{A}{K^2 N^2 r^2} & 0 & 0 \\
        0 & 0 & \frac{1}{K^2 r^2} & 0 \\
        0 & 0 & 0 & e^{-\mu } \\
        \end{array}
        \right),
	\end{equation}
	where is defined as 
	\begin{equation}
	    \label{A2}
	    A=(K r \omega -N \csc (\theta )) (K r \omega +N \csc (\theta )).
	\end{equation}
    This space-time has associated the following tetrad\footnote{We use the same notation as in Ref.~\cite{Chandrasekhar:1985kt}}
    \begin{equation}
        \label{A3}
        \begin{aligned}
            e_{(0)\mu}&=(-N,0,0,0),\\
            e_{(1)\mu}&=(-rK\omega\sin\theta, rK\sin\theta,0,0),\\
            e_{(2)\mu}&=(0,0,rK,0),\\
            e_{(3)\mu}&=(0,0,0,e^\frac{\mu}{2}).
        \end{aligned}
    \end{equation}
    using the relation $e^{\;\;\;\;\mu}_{(a)}=g^{\mu\nu}e_{(a)\nu}$, the contravariant vectors are given by
        \begin{equation}
        \label{A4}
        \begin{aligned}
            e^{\;\;\;\;\mu}_{(0)}&=\left(\frac{1}{N},\frac{\omega}{N},0,0\right)\\
            e^{\;\;\;\;\mu}_{(1)}&=\left(0, \frac{1}{rK\sin\theta},0,0\right),\\
            e^{\;\;\;\;\mu}_{(2)}&=\left(0,0,\frac{1}{rK},0\right),\\
           e^{\;\;\;\;\mu}_{(3)}&=\left(0,0,0,e^{-\frac{\mu}{2}}\right).
        \end{aligned}
    \end{equation}
    Therefore, for the tetrad so defined, we have 
    \begin{equation}
        \label{A5}
        e^{\;\;\;\;\mu}_{(a)}e_{(b)\mu}=\eta_{(a)(b)}=\left(
        \begin{array}{cccc}
        -1 & 0 & 0 & 0\\
         0 & 1 & 0 & 0\\
         0 & 0 & 1 & 0\\
         0 & 0 & 0 & 1
        \end{array}
        \right).
    \end{equation}
    This means that the chosen frame is Minkowskian, i.e. it represents locally an \textit{inertial frame}. It is straightforward to show that the components of the line element (\ref{s2be1}) can be obtained using the relation 
    \begin{equation}
        \label{A6}
        g_{\mu\nu}=\eta^{(a)(b)}e_{(a)\mu}e_{(b)\nu}.
    \end{equation}
    
    The components of the four-velocity are given by 
    \begin{equation}
        \label{A7}
        \begin{array}{ccc}
            u^0=\frac{dt}{d\lambda}, & u^1=\Omega u^0, & u^\alpha=v^{\alpha} u^0\ , 
        \end{array}
    \end{equation}
    where $\alpha=2$, $3$, $v^\alpha=dx^\alpha/dt$, and $\Omega\equiv d\varphi/dt$. In the inertial frame, the components of the four-velocity are computed using the following relation
    \begin{equation}
        \label{A8}
        u^{(a)}=\eta^{(a)(b)}e_{(b)\mu}u^\mu,
    \end{equation}
    from which 
    \begin{equation}
        \label{A9}
        \begin{aligned}
            u^{(0)}&= N u^0,\\
            u^{(1)}&= (\Omega -\omega)rK\sin\theta u^0,\\
            u^{(2)}&= rKv^2u^0,\\
            u^{(3)}&= e^\frac{\mu}{2}v^3u^0.\\
        \end{aligned}
    \end{equation}
    From the second relation in Eq.~(\ref{A9}), we can conclude that a point moving in circular motion with angular velocity $\Omega$ in the coordinate system $(t,\varphi,\theta,r)$, will move with an angular velocity $(\Omega-\omega)rK\sin\theta u^0$ in the inertial frame. Similarly, a point which is considered at rest in the local inertial frame (i.e. $u^{(1)}=u^{(2)}=u^{(3)}=0$), will have an angular velocity $\omega$ in the coordinate frame. Therefore, the non-vanishing of $\omega$ is said to describe a dragging of the inertial frame. Since the space-time is asymptotically flat, then $\omega=2J/r^{3}$. 

%%%%%%%%%%%%%%%%%%%%%%%%%%%%%%%%%%%%%%%%%%%%%%%%%%%%%%%%%%%%%%%%%%%%%%%%%%%%% Appendix B %%%%%%%%%%%%%%%%%%%%%%%%%%%%%%%%%%%
%%%%%%%%%%%%%%%%%%%%%%%%%%%%%%%%%%%%%%%%%%%%%%%%%%%%%%%%%%%%%%       
    %\newpage
    \section{Calculation of $Dp^\alpha/d\lambda$ for $t$, $r$ and $\varphi$\label{appenB}}
    
    From the first MPD equation in Eq.~(\ref{s3e1}), we obtain
    \begin{equation}
        \label{B1}
        \frac{Dp_\nu}{d\lambda}=-\frac{1}{2}R_{\nu\beta\delta\sigma}u^\beta S^{\delta\sigma}.
    \end{equation}
    from which one can compute the $Dp_t/d\lambda$, $Dp_r/d\lambda$, and $Dp_\varphi/d\lambda$ in terms of the components of the Riemann tensor. 
    
    In the case of $Dp_t/d\lambda$, one obtains the following expression,
    \begin{equation}
        \label{B2}
        \frac{Dp_t}{d\lambda}=-\frac{1}{2}\left[2R_{tt\delta\sigma}u^tS^{\delta\sigma}+
        2R_{tr\delta\sigma}u^rS^{\delta\sigma}+2R_{t\varphi\delta\sigma}u^\varphi S^{\delta\sigma}\right].
    \end{equation}
    The factor $2$ comes because $R_{\nu\beta\delta\sigma}$ and $S^{\delta\sigma}$ are skew-symmetric tensors. Therefore, we have to count twice in the sum because $R_{ab\delta\sigma} S^{\delta\sigma}= R_{ab\sigma\delta} S^{\sigma\delta}$. Hence, from Eq.~(\ref{B2}), we obtain
    \begin{equation}
        \label{B3}
        \begin{aligned}
        \frac{Dp_t}{d\lambda}=&-\left(R_{tttr}S^{tr}+R_{ttt\varphi}S^{t\varphi}+R_{ttr\varphi}S^{r\varphi}\right)u^t\\
        &-\left(R_{trtr}S^{tr}+R_{trt\varphi}S^{t\varphi}+R_{trr\varphi}S^{r\varphi}\right)u^r\\
        &-\left(R_{t\varphi t r}S^{tr}+R_{t\varphi t\varphi}S^{t\varphi}+R_{t\varphi r\varphi}S^{r\varphi}\right)u^\varphi.
        \end{aligned}
    \end{equation}
    Then, after using Eq.~(\ref{s3ae3}) and considering the non-vanishing components of the Riemann tensor, the last expression reduces to 
    \begin{equation}
        \label{B4}
        \frac{Dp_t}{d\lambda}=\frac{S^{\varphi r}}{p_
        t}\left[\left(p_\varphi R_{trtr}-p_t R_{tr\varphi r}\right)u^r-p_r R_{t\varphi t\varphi}u^\varphi\right].
    \end{equation}
    \begin{widetext}
    In a similar way, we obtain
    \begin{equation}
        \label{B5}
        \begin{aligned}
        \frac{Dp_r}{d\lambda}&=-\left[R_{rt\delta\sigma}u^tS^{\delta\sigma}+
        R_{rr\delta\sigma}u^rS^{\delta\sigma}+R_{r\varphi\delta\sigma}u^\varphi S^{\delta\sigma}\right],\\\\
        \frac{Dp_\varphi}{d\lambda}&=-\left[R_{\varphi t\delta\sigma}u^tS^{\delta\sigma}+
        R_{\varphi r\delta\sigma}u^rS^{\delta\sigma}+R_{\varphi\varphi\delta\sigma}u^\varphi S^{\delta\sigma}\right].
        \end{aligned}
    \end{equation}
    From which, after using Eq.~(\ref{s3ae3}) and considering the non-vanishing components of the Riemann tensor, we obtain 
    \begin{equation}
        \label{B6}
        \begin{aligned}
        \frac{Dp_r}{d\lambda}&=\frac{S^{\varphi r}}{p_t}\left[(p_\varphi R_{rttr}+p_tR_{rtr\varphi})u^t+(p_\varphi R_{r\varphi tr}+p_t R_{r\varphi r\varphi})u^\varphi\right],\\\\
        \frac{Dp_\varphi}{d\lambda}&=\frac{S^{\varphi r}}{p_t}
        \left[u^r(p_\varphi R_{\varphi rtr}+p_t R_{\varphi rr\varphi})-p_r R_{\varphi tt\varphi}u^t\right].
        \end{aligned}
    \end{equation}
    Note that in the case of a non-rotating wormhole, $R_{tr\varphi r}=0$. Therefore, Eqs.~(\ref{B4}), (\ref{B6}) reduces to the Eqs.~(A11) and (A12) of Ref.~\cite{Benavides-Gallego:2021lqn}, respectively. 
    \end{widetext}
%%%%%%%%%%%%%%%%%%%%%%%%%%%%%%%%%%%%%%%%%%%%%%%%%%%%%%%%%%%%%%%%%%%%%%%%%%%%%%%%%% Bibliography %%%%%%%%%%%%%%%%%%%%%%%%%%%%
%%%%%%%%%%%%%%%%%%%%%%%%%%%%%%%%%%%%%%%%%%%%%%%%%%%%%%%%%%%%%%

	\end{document}